\documentclass{aa}

\def\dg{$^{\circ}$}

\def\rsun{R$_\odot$}
\def\msun{M$_\odot$}
\def\var{OGLE-LMC-ECL-14413}

\usepackage{graphicx}
\usepackage{tabularx}
\usepackage{txfonts}
\usepackage{hyperref}
\begin{document}

  \title{Examining Brightness Variability, Accretion Disk and Evolutionary Stage of Binary 
  OGLE-LMC-ECL-14413  }
   \author{R.E.\,Mennickent
          \inst{1}
                   \and
          G.\,Djura\v{s}evi\'c
          \inst{2,3}
            \and 
             J.A.\,Rosales\inst{1} 
              \and 
          J.\,Garc\'es\inst{1}
          \and
           J.\,Petrovi\'c\inst{2}
                      \and         
          D.R. G. \,Schleicher\inst{1}
          \and
          M.\,Jurkovic\inst{2}
          \and
          I.\,Soszy\'nski\inst{4}
         \and
         J.G.\,Fern\'andez-Trincado\inst{5} 
          }
   \institute{Universidad de Concepci\'on, Departamento de Astronom\'{\i}a, Casilla 160-C, Concepci\'on, Chile\\
              \email{rmennick@udec.cl}
         \and
             Astronomical Observatory, Volgina 7, 11060 Belgrade 38, Serbia
                     \and
                     Issac Newton institute of Chile, Yugoslavia Branch, 11060, Belgrade, Serbia       
                     \and
                     Astronomical Observatory, University of Warsaw, Al. Ujazdowskie 4, 00-478 Warszawa, Poland
                    \and
                    Instituto de Astronom\'ia, Universidad Cat\'olica del Norte, Av. Angamos 0610, Antofagasta, Chile 
             }
   \date{Received XX XX, 2021; accepted XX XX, 2021}

\authorrunning{Mennickent, Djura\v{s}evi\'c, Rosales et al.}
\titlerunning{The binary OGLE-LMC-ECL-14413}

  \abstract
   {    Several intermediate-mass close binary systems exhibit photometric cycles longer than their orbital periods, potentially due to changes in their accretion disks. Past studies indicate that analyzing historical light curves can provide valuable insights into disk evolution and track variations in mass transfer rates within these systems. }
   {Our study aims to elucidate both short-term and long-term variations in the light curve of the eclipsing system \var, with a particular focus on the unusual reversals in eclipse depth. We aim to clarify the role of the accretion disk in these fluctuations, especially in long-cycle changes spanning hundreds of days. Additionally, we seek to determine the evolutionary stage of the system and gain insights into the internal structure of its stellar components.}
   { We analyzed photometric time series from the Optical Gravitational Lensing Experiment (OGLE) project in $I$ and $V$ bands, and from the Massive Compact Halo Objects (MACHO) project in $B_{M}$ and $R_{M}$ bands, covering a period of 30.85 years. Using light curve data from 27 epochs, we constructed models of the accretion disk. An optimized simplex algorithm was employed to solve the inverse problem, deriving the best-fit parameters for the stars, orbit, and disk. We also utilized the Modules for Experiments in Stellar Astrophysics (\texttt{MESA}) software to assess the evolutionary stage of the binary system, investigating the progenitors and potential future developments.} 
   {We found an orbital period of 38.15917 $\pm$ 0.00054 days and a long-term cycle of approximately 780 days. Temperature, mass, radius, and surface gravity values were determined for both stars. The photometric orbital cycle and the long-term cycle are consistent with a disk containing variable physical properties, including two shock regions. The disk encircles the more massive star and the system brightness variations aligns with the long-term cycle at orbital phase 0.25. Our mass transfer rate calculations correspond to these brightness changes. \texttt{MESA} simulations indicate weak magnetic fields in the donor star's subsurface, which are insufficient to influence mass transfer rates significantly.} 
  
   {}


   \maketitle
%

\section{Introduction}

Interacting binary stars constitute complex natural laboratories, allowing the study of a wide range of phenomena such as the physics of accretion disks, stellar winds, gas dynamics of mass flows between stars, angular momentum loss and balance, stellar rotation and the effects of tidal forces. It is believed that a significant portion of stars are members of multiple star systems, and that interactions between closely bound, gravitationally linked stars are common in the Universe. The most energetic phenomena detected so far, such as the merging of black holes or neutron stars, are considered to be the final stages in the evolution of binary systems that have previously undergone phases of mass transfer and angular momentum loss. A good review of binary star evolution is given by \citet{2011epbm.book.....E}.

Among the myriad of interacting binary stars, a particularly intriguing group displays a photometric periodicity longer than the orbital period of typical amplitude at the $I$-band of 0.1 or 0.2 mag, whose nature still defies explanation \citep{Mennickent2003, Mennickent2017}. These are known as Double Periodic Variable stars (DPVs), with masses around 10 solar masses and orbital periods ranging from 3 to 100 days. In these systems, there is invariably a B-type dwarf star, enveloped in a gas disk fed by a cooler giant star of approximately one solar mass that fills its Roche lobe. More than 200 DPVs have been reported in the Galaxy and the Magellanic Clouds \citep{Mennickent2003, 2010AcA....60..179P, Pawlak2013, Mennickent2016, 2021ApJ...922...30R} and the well studied systems reveal higher luminosities and hotter temperatures than normal Algol-type binaries \citep{Mennickent2016}. Some examples of these systems previously detected in our Galaxy include  RX\,Cas \citep{1944ApJ...100..230G}, AU\,Mon \citep{1980A&A....85..342L}, $\beta$ Lyr \citep{1989SSRv...50...35G}, V\,360 Lac \citep{1997A&A...324..965H} and CX\,Dra \citep{1998HvaOB..22...17K}. 

\begin{figure*}
\scalebox{1}[1]{\includegraphics[angle=0,width=18cm]{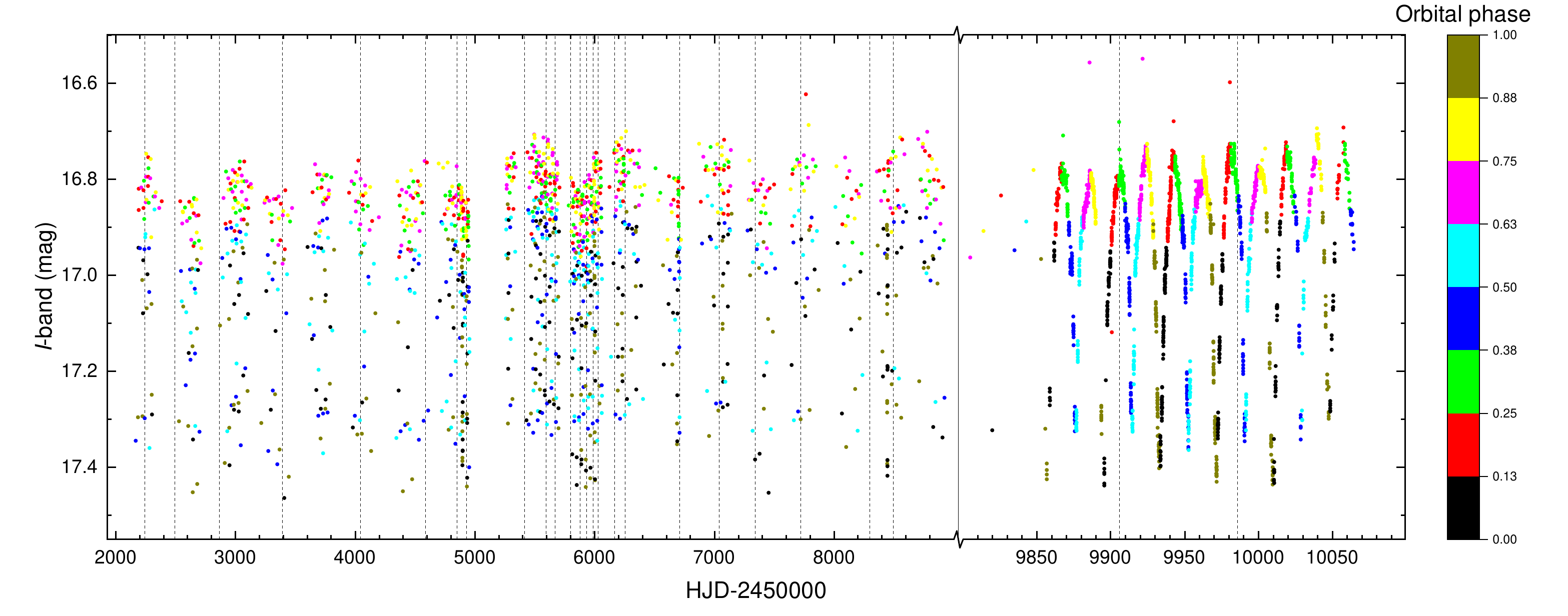}}
\caption{OGLE $I$-band light curve of OGLE-LMC-ECL-14413. Colors indicate different orbital phases and dashed lines show mid times of the data strings listed in Table \ref{tab:strings}.}
\label{fig:Iband}
\end{figure*}

It has been proposed that the longer period of DPVs results from variations in the mass transfer rate, modulated by oscillations in the equatorial radius of the donor star, governed by a magnetic dynamo cycle \citep{2017A&A...602A.109S}. This would induce observable changes in the gas disk over extended time scales. While few DPVs have been investigated with high levels of detail and precision, recent analyses of light curves from the  Optical Gravitational Lensing Experiment (OGLE) database for specific objects have yielded intriguing insights. 

For instance, OGLE-LMC-DPV-097 displays a notable  long-term cycle length of $\sim$ 306 days, with a significant fluctuation of $\sim$ 0.8 magnitudes in the $I$-band. This system orbits every 7\fd75  and comprises stars with modest masses of 5.5 and 1.1\msun. A key finding from examining OGLE-LMC-DPV-097 is the systematic variation in its orbital light curve throughout the  long-term cycle. At the long cycle's nadir, the secondary eclipse is almost imperceptible, and in the rising phase, the system's brightness is more pronounced in the first quadrant compared to the second. These observations can be attributed to the variations in the disk's temperature and dimensions, alongside fluctuations in the luminous and heated spots. The disk's radius contracts to 7.5 \rsun\ at the cycle's low point and expands to 15.3 \rsun\ during the rise, with its outer edge temperature varying from 6870 K to 4030 K between these phases \citep{Garces2018}.

In examining the photometric characteristics of OGLE-BLG-ECL-157529, we encounter another intriguing scenario. This system is marked by orbital and extended periods of 24.8 days and 850 days, respectively, with stellar masses similar to OGLE-LMC-DPV-097. It is observed that while the primary minimum's magnitude remains largely stable, the secondary minimum exhibits significant fluctuations. The secondary minimum occurs due to the eclipse of the donor star by the combined structure of the accretor and its disk. This suggests the presence of a fluctuating accretion disk. Notably, the brightness at the orbital phase of 0.25 aligns closely with the  long-term cycle. The disk's fractional radius (F$_{d}$), a metric for the accretor's Roche lobe occupancy, reveals that the disk expands beyond the tidal radius during peak activity and longer cycles. Conversely, F$_{d}$ decreases below the tidal radius during quicker, less intense  long-term cycles in later epochs. In this system, the  long-term cycle's minimum is attributed to a thicker disk that obscures more of the accretor, with a higher mass transfer rate resulting in a hotter disk at the outer edge. The disk's radius oscillations over hundreds of days defy explanation through viscous energy release, as seen in the superhumps of precessing disks in cataclysmic variables like SU UMa stars. This is because Lindblad resonance regions lie well outside the disk's radius. As an alternative, the hypothesis of variable mass injection as a cause for long-term photometric variability has been suggested \citep{2021A&A...653A..89M}.

\begin{table}
\centering
\caption{Photometric observations.}
\tiny
\label{tab:observinglog}
\begin{tabular}{lrrrr} 
\hline
\small
band &  N      & $\rm{HJD'_{start}}$      &  $\rm{HJD'_{end}}$   & mean $\pm$ std (mag) \\
\hline
$B_{M}$&1315&-1173.8652 & 1541.9973 &-5.850 $\pm$ 0.175\\
$R_{M}$&1510&-1173.8652 & 1541.9973 &-6.417 $\pm$ 0.173\\
$I$ &535 &2167.8681&4955.5136&16.985  $\pm$ 0.180 \\
$V$ &136 &2994.7006 &4955.4684& 18.143 $\pm$ 0.180 \\
$I$  & 3364     &   5260.6121   &  10064.5314  &16.946   $\pm$ 0.182 \\
$V$   & 280     &   5260.6596  &  10061.4713 &18.111 $\pm$ 0.178    \\
\hline
\vspace{0.05cm}
\end{tabular} \\

Note: Summary of OGLE-III, OGLE-IV and MACHO photometric observations  analyzed in this paper. The number of measurements, starting and ending times, and average magnitude and their standard deviation are given. 
HJD' = HJD-2450000. The uncertainty of a single OGLE measurement varies between  4 and 6 mmag and for MACHO $R_{M}$ and $B_{M}$  are typically 0.048 mag and 0.058 mag.
\end{table}

\begin{figure}
\scalebox{1}[1]{\includegraphics[angle=0,width=8cm]{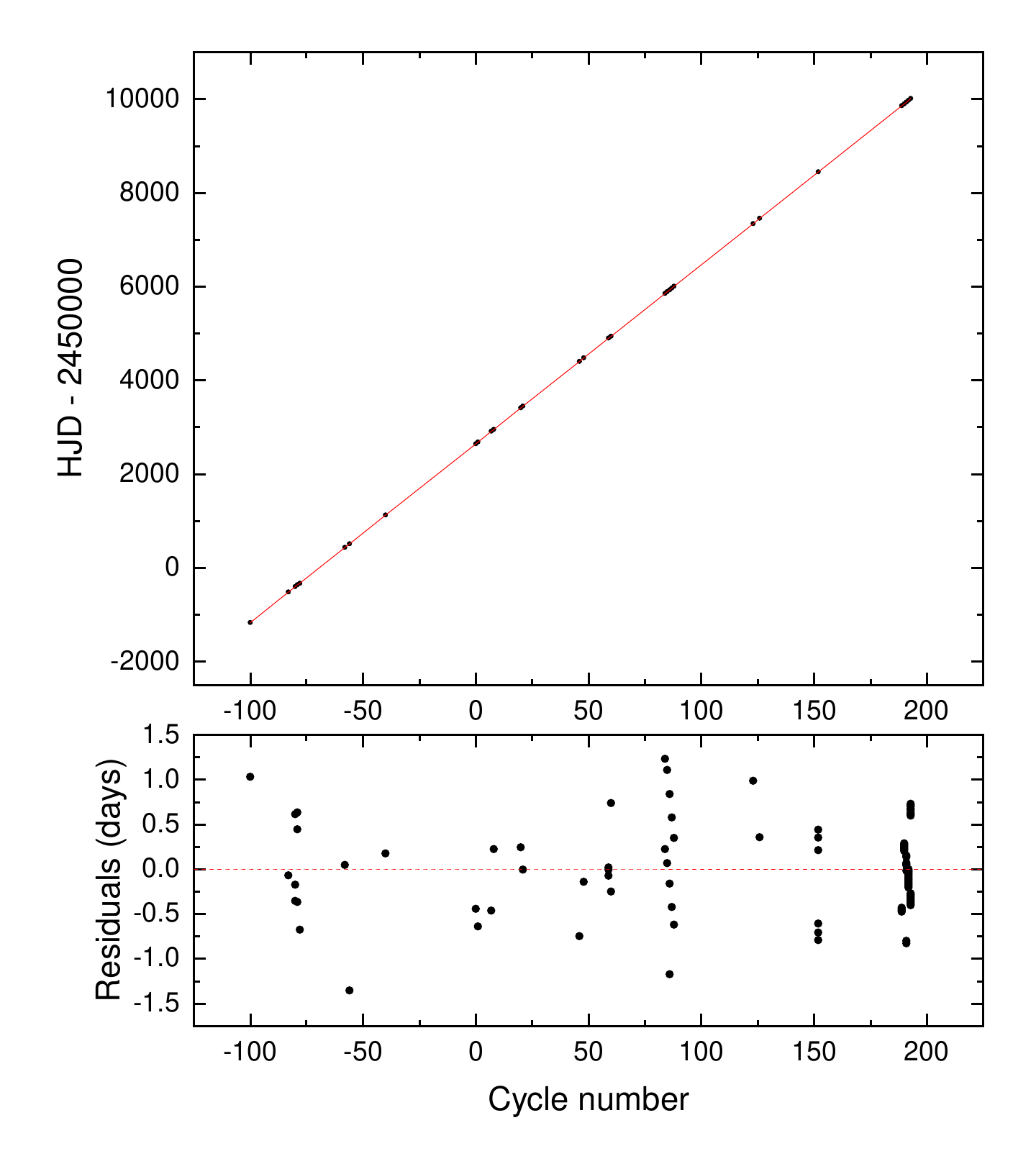}}
\caption{Times of main minima versus cycle number and the best linear fit along with residuals. The slope of the fit gives the orbital period, viz. P$_{\rm{o}}$ = 38\fd15917 $\pm$ 0\fd00054.}
\label{fig:period}
\end{figure}

\begin{figure*}
\scalebox{1}[1]{\includegraphics[angle=0,width=18cm]{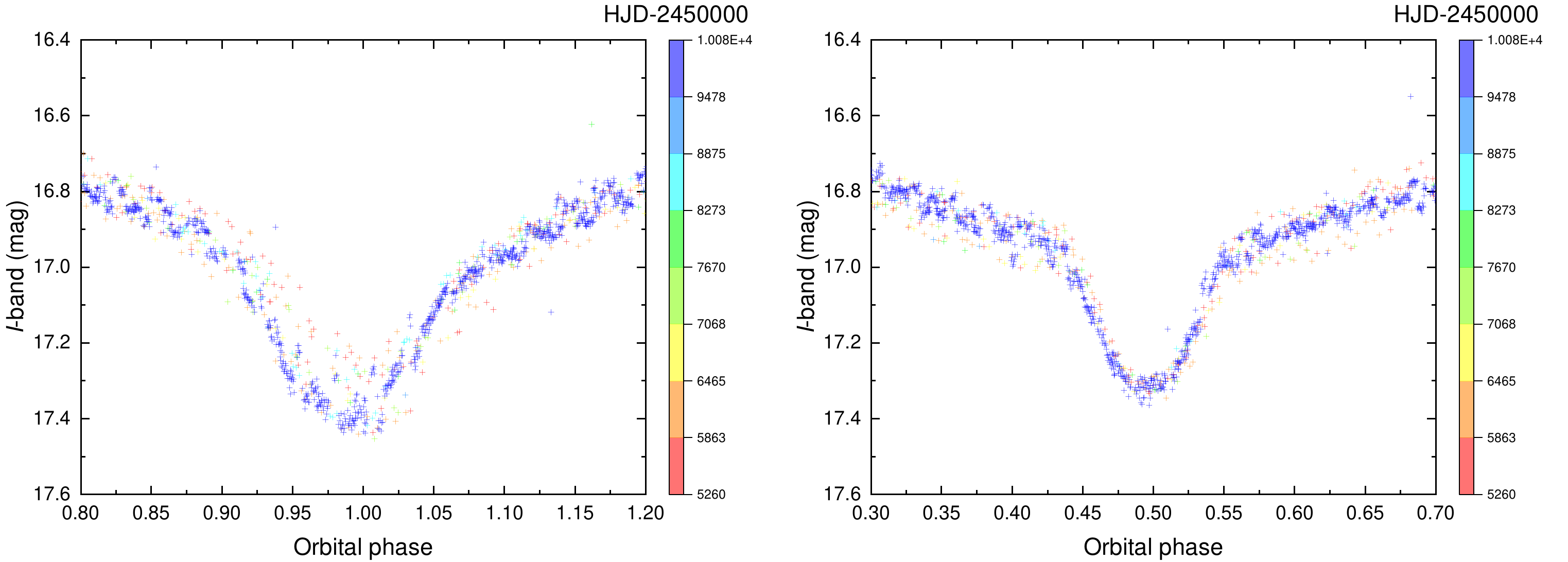}}
\caption{OGLE $I$-band light curve of OGLE-LMC-ECL-14413 phased with the orbital period around the eclipse of the donor (left) and the gainer (right). Colors indicate different HJDs.}
\label{fig:eclipses}
\end{figure*}

In order to complement the above studies, and provide a much complete  picture of the behavior of these DPV systems, we present a photometric study of the eclipsing binary DPV \var\ \citep{2011AcA....61..103G, 2016AcA....66..421P}. This binary is also know as OGLE-III LMC162.7.89414, OGLE-IV LMC503.02.14123, TIC 373520393, MACHO 78.6943.2764 and 2MASS J05232086-6950586 and their data include $\alpha_{2000}$=05:23:20.96, $\delta_{2000}$=$-$69:50:58.0\footnote{https://gea.esac.esa.int/archive/} and $I$= 16.817 mag and $V$= 18.001 mag\footnote{https://simbad.cfa.harvard.edu/simbad/}. 
This object was selected because of its relatively long orbital period and peculiar characteristics.
The OGLE Collection of Variable Stars Website indicates that the system has an orbital period of 38\fd1613814. The system is mentioned as a unique  DPV  among the sample of LMC DPVs studied by \citet{2010AcA....60..179P}. These authors noticed  that apparently the amplitude of the  long-term cycle gets smaller during one of the eclipses, also that the depth of the usually deeper minimum changes from cycle to cycle. They argue that a similar depth of both minima suggests that temperatures of both components are similar and conclude, from the observation that the mean light curve is typical for semi-detached or contact binary systems, that the radius of the donor star is one of the biggest among donors of known DPVs, or the total mass of the binary is smaller than for the other DPVs. The authors also argue that the donor eclipses the disk and the gainer during the shallower and more stable eclipse.

\begin{figure}
\begin{center}
\scalebox{1}[1]{\includegraphics[angle=0,width=7cm]{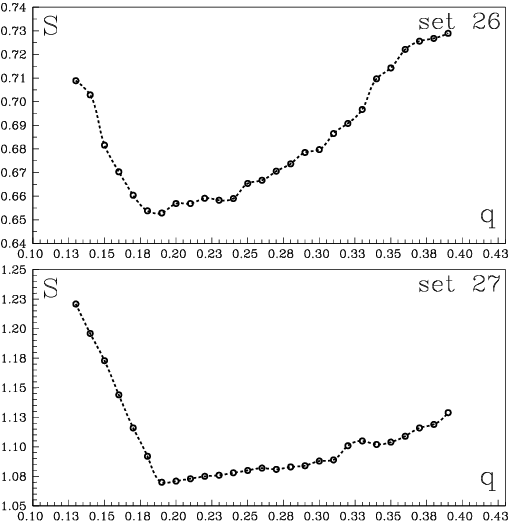}}
\caption{Parameter S = $\Sigma$ (O-C)$^{2}$ for the fits done to the light curve of datasets 26 and 27, as a function of mass ratio.  }
\label{fig:qsearch}
\end{center}
\end{figure}

\begin{figure}
\scalebox{1}[1]{\includegraphics[angle=0,width=8.5cm]{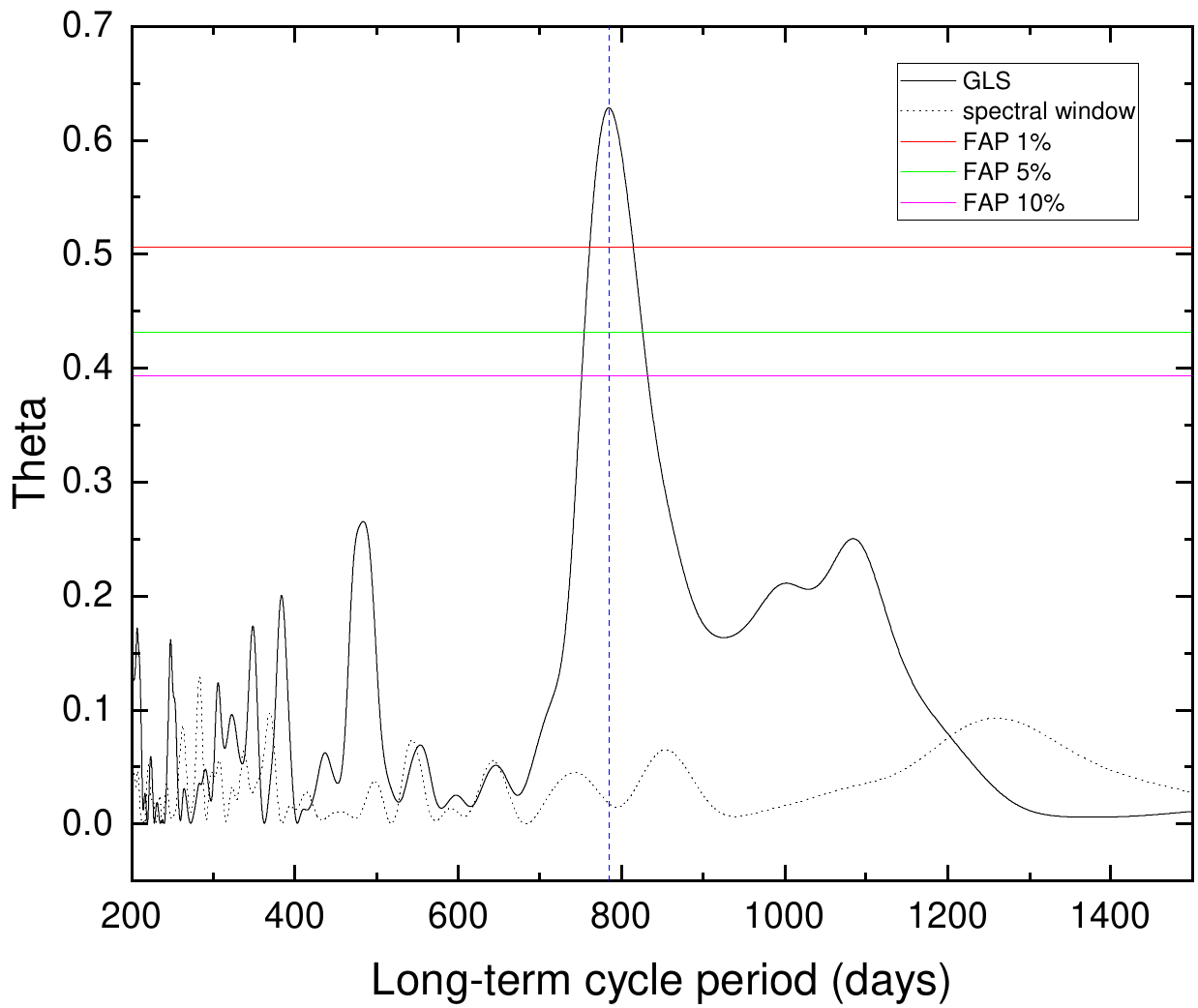}}
\caption{Generalized Lomb Scargle periodogram for the magnitudes measured at phase 0.25. The vertical line shows the maximum value at 784\fd8. False alarm probabilities and the spectral window are also shown.}
\label{fig:plotGLS}
\end{figure}

 \begin{figure}
\scalebox{1}[1]{\includegraphics[angle=0,width=8.5cm]{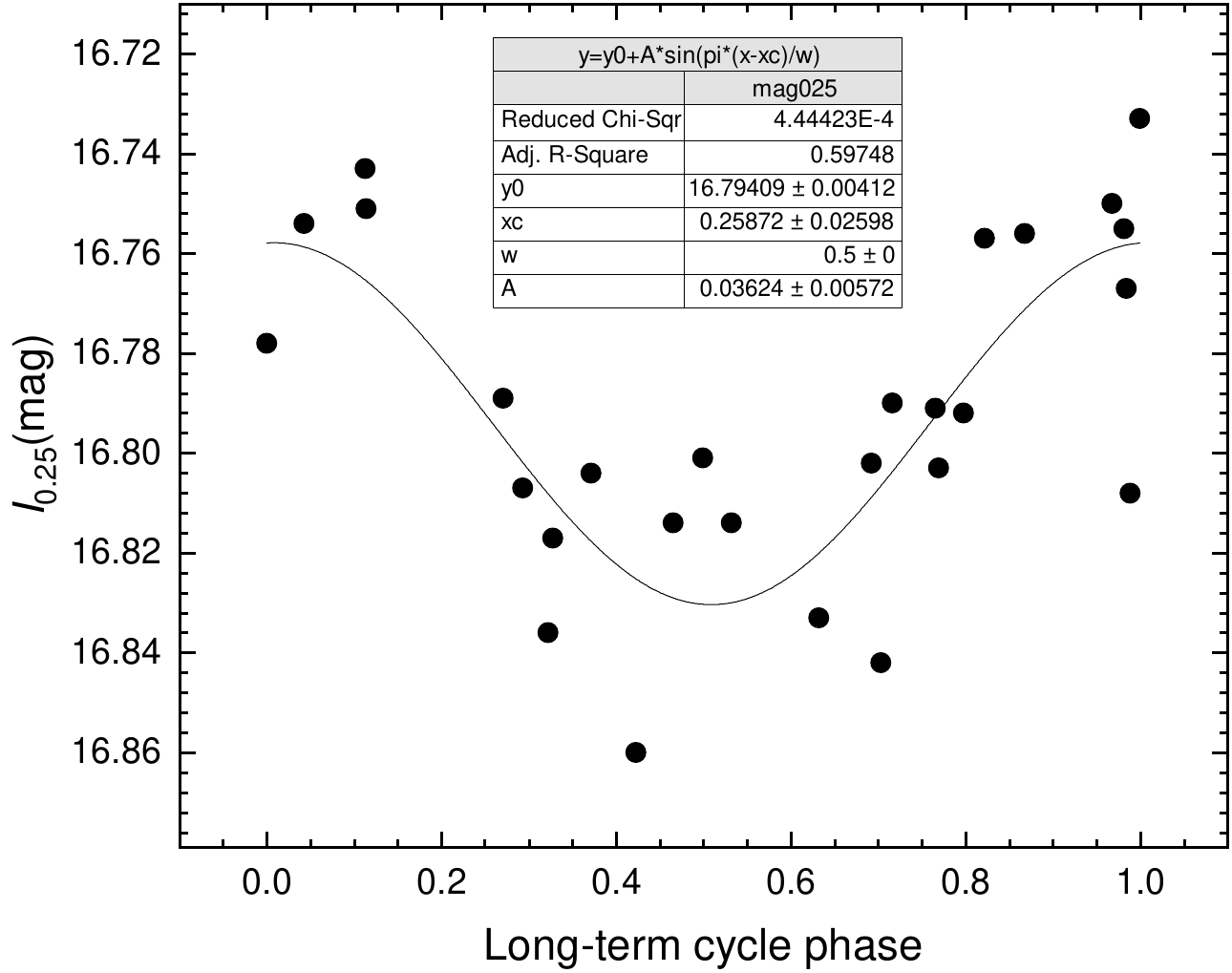}}
\scalebox{1}[1]{\includegraphics[angle=0,width=8.5cm]{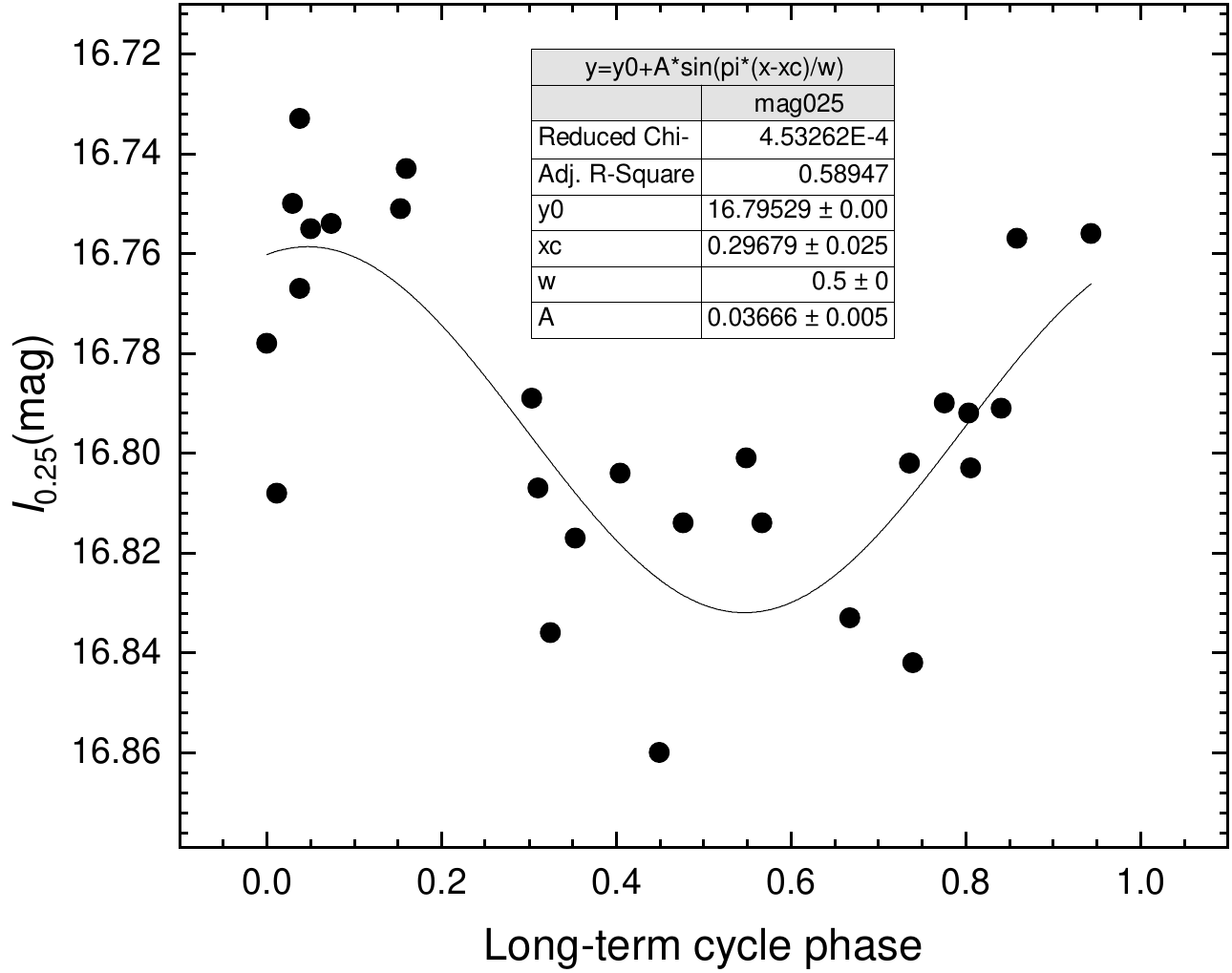}}
\caption{Magnitudes at phase 0.25 phased with the period of 784\fd8 (up) and 778\fd8 (down). Phase zero corresponds to HJD' = 2241.7735. Typical error is 0.005 mag. The best sinus fit are also shown.}
\label{fig:phased}
\end{figure}

 \begin{figure}
 \begin{center}
\scalebox{1}[1]{\includegraphics[angle=0,width=8.5cm]{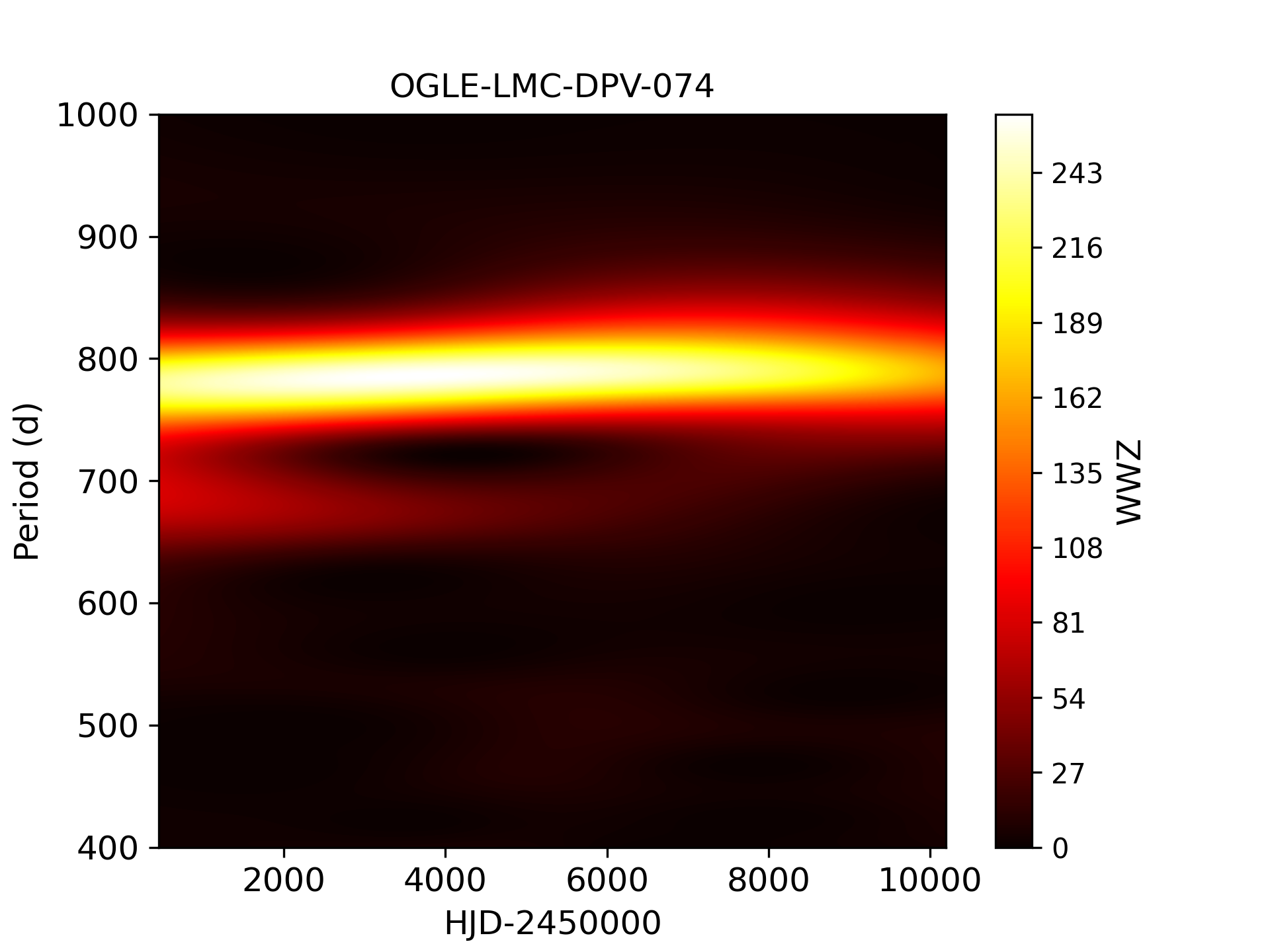}}
\caption{WWZ transform shows a strong signal around 778.8 days.}
\label{fig:wwz}
\end{center}
\end{figure}

\section{Photometric data and light curve}

The photometric time series analyzed in this study consists of 3899 $I$-band data points and 416 $V$-band datapoints obtained by The 
Optical Gravitational Lensing Experiment \citep[OGLE,][]{2015AcA....65....1U}. The $I$-band light curve shows the typical variability of an eclipsing binary, plus  a long-term variability that is remarkable in the upper boundary of the dataset, that is shaped by the brighter datapoints (Fig. \ref{fig:Iband}). In addition, we study 1315 $B_{M}$ and 1510 $R_{M}$ data points obtained  by the Massive Compact Halo Objects (MACHO) project\footnote{https://datacommons.anu.edu.au/DataCommons/item/anudc:3255}. 
The whole dataset, summarized in Table \ref{tab:observinglog}, spans a time interval of  30.85 yr.

In this paper, as a starting point,  we use the ephemeris for the occultation of the cooler star,  provided by the OGLE team \citep[][no error is given]{2016AcA....66..421P}:
\small
\begin{eqnarray}
\rm HJD&=&\rm 245\,7033\fd527 + \rm 38\fd1613814 \, E.
\end{eqnarray}
\normalsize

\noindent  When comparing DPV systems that feature alternating depths of primary and secondary minima, it is crucial to maintain a consistent physical definition of these eclipses. As mentioned earlier, we have adhered to the methodology established by  \citet[][]{2016AcA....66..421P}. 

In order to check the constancy of the orbital period, we performed an analysis of the
times of the primary eclipse using the method of comparison of predicted times with
observed times of eclipses, described
by \citet{2005ASPC..335....3S}. As eclipse timings we have chosen the fainter points
in the colored data strings shown in Fig. \ref{fig:Iband}. 
We find that the data is compatible with a constant orbital period and there is no practical difference with the ephemerides provided by the OGLE team.

Furthermore, in order to improve the orbital period, we performed an analysis of the main eclipse times. For that, since most eclipses are not well defined individually, and after a visual inspection of the light curve, we selected  candidates for minima as those observations with magnitudes at bands $I$, $B_{M}$ and $R_{M}$ fainter than 17.4, -5.35 and -6.00 and few outliers with phases discordant were rejected. The 95 eclipse times thus selected are shown in Table \ref{tab:minimatimes}. 

Using the published orbital period as a trial, we calculated the cycle numbers for every minimum and fit a straight among the (N, time) pairs obtaining the following improved ephemeris for the main eclipse (Fig. \ref{fig:period}):

\small
\begin{eqnarray}
\rm HJD&=&\rm 245\,2645\fd198(83) + \rm 38\fd15917(54)  \, E.
\end{eqnarray}
\normalsize

\noindent
This new orbital period has a difference of 0\fd00221 with the previously reported period, this accumulates 0\fd66 during the 300 cycles of observations, or equivalently 0.017 orbital phase units. We also calculated the orbital period for the OGLE data yielding  P$_{\rm o}$ = 38\fd159 $\pm$ 0\fd012  using the Phase Dispersion Minimization software \citep[\texttt{PDM,}][]{1978ApJ...224..953S} and  P$_{\rm o}$ = 38\fd123 $\pm$ 0\fd095 using the Generalized Lomb Scargle software \citep[\texttt{GLS,}][]{2009A&A...496..577Z}.

The light curve variability cannot be understood in terms of a   two stars system, and the presence of circumstellar matter around the gainer is suggested by the observations (Fig. \ref{fig:eclipses}). It is interesting that the   eclipse  observed at phase 0.0, that sometimes is the deeper one, corresponds to occultation of the donor, since it is the eclipse with larger residuals and inter-epoch variability. The cooler star (the donor) is eclipsed by a variable physical structure. However, the depth of eclipse at phase 0.5 is rather stable, compatible with the stability of the donor and the occultation of the whole disk or a large fraction  of it.  The roughly constant shape of this eclipse confirms our ephemerides. Furthermore, there is no evidence for changes in the orbital period. The light curve outside the eclipse of the gainer, shows larger scatter before epochs at HJD' $\sim$ 8273. On the other hand, at some epochs, the ingress of the eclipse of the donor is delayed, producing an eclipse of asymmetrical shape.  These aspects could be attributed to variable amounts of circumstellar mass  in the line of sight, considering the interacting nature of this binary discussed later in this paper.

While the $I$-band light curve is dense enough to allow a good analysis of orbital and system parameters, the $V$-band data are scarce, and we used them only for determination of the color at minimum and temperature of the donor star. During the secondary eclipse, the color $V-I$ = 1.263 mag. If we take 
the mean $E(V-I)$ = 0.113 $\pm$ 0.060 mag for the LMC \citep{2019A&A...628A..51J}, then we get $(V-I)_0$= 1.150  $\pm$ 0.06 mag, implying $T_2$= 5100 K, according to the calibration given by  \citet{2020A&A...641A..91M}.  As a reference, the OGLE database provides $V-I$ = 1.156 mag, suggesting $T_2$= 4982 K, but this value rests on the average color from the entire light curve. 
We can also use the reddening maps by \citet{2021ApJS..252...23S} to get $E(V-I)$ = 0.106 $\pm$  0.086 which gives $(V-I)_0$= 1.157  $\pm$ 0.086 mag. 
We decided to keep fixed the temperature of the  donor in $T_2$ = 5000  with an uncertainty of  $\pm$ 200 K, neglecting errors that could come from the disk contribution to the color at the secondary eclipse.

\section{Methodology and light curve analysis}

In this section we discuss our model for the light curve based only on OGLE data, since the errors associated to MACHO photometry are 10 times larges than those for OGLE data. Therefore, MACHO light curve were used to find additional times of minima only. In order to study the photometric changes, we divided the whole dataset in subsamples of 27 consecutive data  sets (Table \ref{tab:strings}).  This allowed us to work with clean orbital light curves with no much influence of the  long-term cycle. The choice of a different number of  segments does not change the results of this paper. We must keep in mind that 
the time range must be large enough to include data to represent adequately the orbital light curve, especially eclipses and quadratures,  but also short enough to exclude as much as possible the variability  due to the  long-term cycle.

\subsection{The light curve model}

Utilizing an advanced simplex algorithm  \citep{DT91}, the inverse problem was tackled by calibrating the light curve to  match the most accurate parameters of the stars-orbit-disk combination for the observed system.  The core principles  of this model and the procedure for synthesizing light curves are well-documented in  the literature \citep{1992Ap&SS.196..267D, 1996Ap&SS.240..317D}. Further developments and enhancements of this model are also reported \citep{2008AJ....136..767D} and have been successfully applied in the analysis of numerous close binary systems \citep[e.g.][]{2013MNRAS.432..799M, 2018MNRAS.476.3039R, 2020A&A...642A.211M}.

The model quantifies the  flux from the binary system as the cumulative output of individual stellar fluxes, coupled with the radiation emanating from an optically  thick accretion disk encircling the more luminous star. It incorporates the spatial effects induced by the observer's viewing angle. The light contribution from the disk is estimated using localized Planck radiation functions at defined temperatures, deliberately omitting the complex calculations involved in radiative transfer. Nevertheless, the model takes into account simpler phenomena like reflection, limb darkening, and gravitational dimming.

The accretion disk is characterized by parameters such as its radius R$_d$, vertical thickness at its central and outer edges (d$_c$ and d$_e$, respectively), and a temperature profile that varies with radial distance  (e.g. \citet{2021A&A...645A..51B}):

\begin{equation}
T(r) = T_{d} \left(\frac{R_{d}}{r}\right)^{a_{T}}, 
\end{equation}

\noindent where T$_d$ signifies the temperature at the disk's outer rim (r = R$_d$)  and a$_T$ is the temperature gradient exponent, constrained to a$_T$ $\leq$ 0.75. 
The exponent a$_T$  indicates the degree to which the radial temperature profile approaches a steady-state configuration (a$_T$ = 0.75).  This  ad-hoc radial temperature dependency implies that the disk's surface is warmer closer to the center and gradually cools with increasing radial distance.

Additionally, the model postulates the presence of two shock regions at the disk's periphery: a \textquoteleft{hot spot\textquoteright} near the conjectured impact point where the gas stream from the inner Lagrangian point collides with the disk, and a \textquoteleft{bright spot\textquoteright}, located elsewhere along the disk's edge. These regions, characterized by variable thickness and temperature compared to the rest of the disk, have been identified in Doppler maps of Algol systems and in hydrodynamic simulations of gas flows in close binary stars \citep[e.g.][]{1996ApJ...459L..99A, 2000A&A...353.1009B, 2012ApJ...760..134A}. Their existence was inferred from the analysis of the orbital light curve of the DPV $\beta$ Lyrae \citep{2013MNRAS.432..799M}, with interferometric and polarimetric studies further substantiating the presence of the hot spot in this system \citep{2012ApJ...750...59L, 2018A&A...618A.112M}. While the genesis of the hot spot is relatively clear, the bright spot's origin could stem from vertical oscillations of gas at the disk's outer edge due to interaction with the incoming matter stream \citep{2017ARep...61..639K}. The model's depiction of a singular hot spot on the disk's edge falls short of accurately representing the intricate and variable nature of the light curves, necessitating the inclusion of the bright spot for a more comprehensive description.

In this model, both the hot and bright spots are defined by their relative temperatures  $A_{hs} \equiv T_{hs}/T_d$ and $A_{bs} \equiv T_{bs}/T_d$, for their angular dimensions $\theta_{hs}$ and $\theta_{bs}$ and their angular locations measured from the line joining the stars in the direction of the orbital motion $\lambda_{hs}$ and   $\lambda_{bs}$. Additionally, $\theta_{rad}$   represents the angle between the line perpendicular to the local disk edge surface and the direction of maximum radiation from the hot spot.
The disk radius is also gauged by the parameter $F_d$ $\equiv$ R$_d$/R$_{yk}$, where  R$_{yk}$ is defined as per \citet{1992Ap&SS.196..267D} as the distance from the center of the hotter star to its Roche lobe, measured perpendicular to the line joining the star centers. In this framework, the disk maintains stability only within the boundary $F_d$ $\leq$ 1, with any material beyond the Roche limit potentially escaping the system or forming a circumbinary envelope, particularly around the Lagrange equilibrium point L$_3$.
The stars' gravity-darkening coefficients are set at $\beta_1$ = 0.25 and $\beta_2$ = 0.08, with albedo coefficients ${\rm A_1= 1.0}$ and ${\rm A_2= 0.5}$,  in accordance with von Zeipel's law for radiative shells and complete re-radiation \citep{1924MNRAS..84..702V}. The limb-darkening for the components was calculated in the way described by \citet{2010MNRAS.409..329D}.

\subsection{Results of the light curve model}

In studies of over-contact or semi-detached binaries using photometric time series devoid of spectroscopic data, the "q-search" method is commonly employed to deduce the system's mass ratio, and it is specially precise for eclipsing systems  \citep{2005Ap&SS.296..221T}.
Applying this technique, convergent solutions were obtained for a range of mass ratios $q = M_2/M_1$, where the subscripts "2" and "1" denote the cooler and hotter stars, respectively.  
We choose datasets 26 and 27 since they accurately represent the orbital light curve with minimum  brightness depressions caused by the disk obscuring the primary star.  We performed the search in the mass ratio interval 0.13-0.40 with a step of 0.01 and found that a mass ratio of 0.19 aligns pretty well with these datasets (Fig. \ref{fig:qsearch} and Fig. \ref{fig:figa3}). 
The $q$ error is obtained at the minimum of the $S$ value of the dataset 26. This dataset has a more symmetrical $S$ curve than dataset 27. The 5\% $S$ increase from the bottom determines a variation of plus minus 0.01, so we choose a conservative value of $\pm$ 0.02 for the mass ratio uncertainty. The $q$ value corroborates earlier findings indicating an average $q$ = 0.23 $\pm$ 0.05 (std) in previous studies of DPVs \citep{Mennickent2016}.  
In this paper we used $f_g$ = 84.1 for the gainer and $\Omega_1$ and $\Omega_2$ equal to  30.702 and 2.208 for the potentials at the gainer and donor surfaces. For that we assumed synchronous rotation of the donor and non-synchronous, critical rotation for the gainer, that fills its critical oval. 
 A second possible model without an accretion disk was not considered, because it leads to an over-contact configuration and requires approximately the same temperatures of the components, which is not in favor of the observations. In addition, the shape of the light-curves for this system is typical for systems with an accretion disk, as well as its changes over time. So far, we have analyzed several similar systems \citep[e.g.][]{2023A&A...670A..94R, 2022A&A...666A..51M}. More reason to reject the model without an accretion disc is our experience with the RY Sct system, which we first modeled within such a model, only to later reject this model based on spectroscopic data and re-analyze the observations within the model with an accretion disc \citep{2001A&A...374..638D, 2008AJ....136..767D}.

These datasets were also instrumental in determining the main stellar and orbital parameters.   For that,  we estimated the mass of the donor star, viz. 1.1 \msun, by assuming a temperature of $T_2 = 5000$ K and applying the corresponding relationships from the Lang tables \citep{1993Ast....21...95L} that indicate a spectral type G5 for a luminosity class III star. Utilizing the derived mass ratio, we then calculated the mass of the primary star, that accounts to 5.8 \msun. Additionally, we employed Kepler's third law to determine the orbital separation between the two stars, 90.73 \rsun. All derived stellar and orbital parameters, along with their errors,  are shown in Table \ref{tab:system}. Estimates  
for the errors of mass, radius, surface gravity and orbital separation are given in the Appendix.

 We keep constant these stellar and orbital parameters for other datasets, focusing the analysis on the variability of disk parameters.
The outcomes of the light curve modeling are detailed in Table \ref{tab:fitpar}.  Uncertainties were estimated with numerical experiments around the best fitting values. They must be observed as generic ones due to the non-homogenous character of the data. Single orbital light curve fittings are presented in \href{https://doi.org/10.5281/zenodo.14192345}{Figs.  A1-A3}, highlighting measurement uncertainties, typical orbital variations, long-term variability, fitting quality, and the system's visual representation. Observed discrepancies exceeding individual data point errors could be attributed to underestimation of formal errors or unaccounted variabilities in the model.  For instance, a possible additional light source is not considered. Another possible cause for these discrepancies could be the span of individual light curves over multiple orbital cycles, introducing long-term variability into the analysis. To address this, a "sliding" polynomial of lower degree was fitted through each light curve segment, with magnitudes at orbital phases 0.25 and 0.75 measured using the corresponding orbital phases and fit values (Table \ref{tab:strings}).

We find that the disk radius ranges from 31.97 \rsun\ to 43.81 \rsun whereas the disk temperature ranges from 3258 K to 4217 K. The disk outer vertical thickness span from 6.1 to 12.9 \rsun\ while the disk height at the inner boundary span from 3.1 to 5.0 \rsun. The hot spot temperature is at average 1.28 $\pm$ 0.08 times higher than the surrounding disk and the bright spot is 1.14 $\pm$ 0.04 times higher than the surrounding disk.  The hot and bright spots are located at 327\fdg2 $\pm$ 12\fdg1 and 94\fdg5 $\pm$ 33\fdg3, respectively, as measured from the line joining the stars, from the donor to the gainer, and in the sense of the binary motion. The hot and bright spots span arcs of 18\fdg6 $\pm$ 3\fdg1 and 43\fdg0 $\pm$ 5\fdg5, respectively, in the disk's outer edge. In general, we do not observe correlations between the fit parameters except for few cases.

\begin{table}
\caption{Data strings.}
\label{tab:strings}
\centering
\small
\begin{tabular}{r c c c c  c c }
\hline
set& HJD'-mid &$\Delta t$ (d) & mag025 &  mag075 & $dM/dt$ & $\Phi_l$\\
\hline
01 &2241.7735&148&16.778&16.760&1.77&0.95  \\ 
02 &2494.6013&340&16.836&16.845&1.38&0.27  \\
03 &2867.6482&401&16.792&16.790&1.90&0.75  \\
04 &3391.7015&641&16.814&16.813&1.46&0.43  \\
05 &4041.3132&653&16.807&16.803&2.00&0.26  \\
06 &4587.2472&433&16.808&16.780&2.72&0.96  \\
07 &4853.1629&95&16.817&16.818  &1.43&0.30   \\
08 &4928.0578&55&16.860&16.851 & 1.22&0.40  \\
09 &5414.5181   &273 & 16.754    &16.734 &1.85 &0.02     \\
10 &5593.4558    &118 & 16.789    &16.734&1.43 &0.25    \\
11 &5672.2730    &38 & 16.804    &16.760  &1.47 &0.35    \\
12 &5798.5795    &151 & 16.814    &16.820&1.08 &0.52  \\
13 &  5877.0519  &58 &  16.833   & 16.837& 1.40 &0.62 \\
14 & 5932.9618  &55  & 16.842    &16.800 & 3.71 &0.69 \\
15 & 5984.5826   &40 & 16.803    &16.763 & 3.33 &0.76 \\
16 & 6026.0065   &59 &  16.757   & 16.770& 3.04 &0.81 \\
17 & 6165.3956   & 144&  16.733  &16.728&2.31  &0.99 \\
18 &  6254.9853 & 133&  16.751  & 16.728 &2.09 &0.10 \\
19 &  6708.9257 &  310&  16.802  & 16.752 & 1.00&0.69  \\
20 &  7038.9236 &  180&  16.743  & 16.721 & 1.92&0.11 \\ 
21 &   7342.2775 & 532&  16.801 &  16.770& 1.80 &0.50  \\
22 &  7722.7651 & 148&   16.767 &  16.735& 2.08 &0.99  \\
23 &  8297.5512 &  369&   16.790 & 16.764& 2.17 &0.73  \\
24 &   8495.0520 & 157&  16.750 &  16.730& 3.24 &0.98 \\
25 &   9290.3548 &1160&    16.755& 16.725& 2.19&0.00 \\
26 &  9905.8385 &77&     16.791 &  16.772& 2.90  &0.79 \\
27 &   9985.8180 & 125&    16.756 &  16.782&2.71&0.89  \\
 \hline      
 \vspace{0.05cm}
\end{tabular} \\
Note: Mid of the HJD' range, the range of days and the magnitudes at orbital phases 0.25 and 0.75 are given, according to the light curve model. Mass transfer rates, normalized to the value of string 19 and discussed in Section 3.3, are also given, along with the phases of the  long-term cycle $\Phi_l$ according to Eq.\,5.

\end{table}

\begin{table}
\centering
\caption{Calculated stellar and orbital parameters.} 
\label{tab:system}
\small 
\begin{tabular}{lclc} 
\hline
 $M_1$  (M$_{\odot}$) & 5.8  $\pm$  0.3 & log \ $g_1$    & 4.08  $\pm$  0.04 \\
 $M_2$  (M$_{\odot}$) & 1.1  $\pm$  0.1 & log \ $g_2$    & 1.78  $\pm$  0.03 \\
 $R_1$ (R$_{\odot}$)  & 3.6  $\pm$  0.1 &$P_{\rm{o}}$  (d) &38.15917 $\pm$ 0.00054  \\
 $R_2$ (R$_{\odot}$)  & 22.4  $\pm$  0.8 &   $a_{\rm{orb}}$ (R$_{\odot}$)    & 90.7 $\pm$  1.8    \\
 $T_1$  (K) &    18701 $\pm$ 208     & $i$ ($^{\rm{o}}$) &     85.8  $\pm$ 0.3\\
  $T_2$  (K)  & 5000  $\pm$ 200  (fixed)       & & \\     
  \hline
   \vspace{0.05cm}
\end{tabular}\\
Note: We include the orbital separation $a_{\rm{orb}}$ and the system inclination $i$.   Indexes 1 and 2 refer to hot and cool stellar components. The errors for masses, radius, log\,g  and orbital separation are those derived in the Appendix, for the other parameters are from light curve fitting.
\end{table}
\normalsize

\subsection{A simple model for the mass transfer rate}

We also calculated an approximation for the mass transfer rate $\dot{M}$, using the following prescription \citep{2021A&A...653A..89M}: 

 \begin{eqnarray}
 \frac{\dot{M}_{2,f}}{\dot{M}_{2,i}}= \frac{  {{R}^2_{disk,f}} [A_{hs,f}T_{disk,f}]^{4} d_{e,f}  \theta_{hs,f} }{{{R}^2_{disk,i}} [A_{hs,i}T_{disk,i}]^{4} d_{e,i} \theta_{hs,i}}.
\end{eqnarray} 

\noindent This equation allows us to estimate relative mass transfer rates at different epochs $i$ and $f$, for a given system. In order to normalize our determined values of $\dot{M}$, we used as reference its minimum value, attained at  the orbital cycle represented by  string 19.  These values are  shown in Table \ref{tab:strings},   and an uncertainty of up to 40\% is derived from the classical formula of error propagation.  The mass transfer rate hence calculated has a mean value of 2.06 with a standard deviation of 0.73, and maximum and minimum values of
3.71 and 1.

\subsection{The  long-term cycle}

In order to investigate the  long-term cycle length, we searched for periodicities in the data of magnitudes at phase 0.25 with the aid of the \texttt{GLS} periodogram. We searched between 200 and 1500 days, and find a prominent peak at 784\fd8 $\pm$ 24\fd3 (Fig. \ref{fig:plotGLS}).
The magnitudes at phase 0.25 follows a long-term tendency. A sinus fit provides overall $I$-band amplitude of 0.036 $\pm$ 0.006  mag although the peak to peak distance is 0.107 mag (Fig. \ref{fig:phased}). 

The  long-term cycle is also revealed in the Weighted Wavelet Z Transform
(WWZ) as defined by \citet{1996AJ....112.1709F}. The WWZ works in a
similar way to the Lomb-Scargle periodogram providing information
about the periods of the signal and the time associated to
those periods. It is very suitable for the analysis of non-stationary
signals and has advantages for the analysis of time-frequency
local characteristics. We notice that the WWZ transform for the $I$-band time series
suggests a stable  long-term cycle length of 780 days, confirming our finding in the periodogram constructed with magnitudes taken around orbital phase 0.25 (Fig. \ref{fig:wwz}).
The period obtained every step of 100 days with the above tool shows a distribution with a mean of 778\fd8 $\pm$ 4\fd6 (std). 
A sinus fit provides overall $I$-band amplitude of 0.037 $\pm$ 0.006  mag and basically the same chi square that the case for the period 784\fd8 (Fig. \ref{fig:phased}).

In the following, and based on the trial ephemerides shown in Fig. \ref{fig:phased}, we use as ephemeris for the maximum of the  long-term cycle: 

\small
\begin{eqnarray}
\rm HJD&=&\rm 245\,2280.7  \pm 218.1 + \rm 778\fd8 \pm 4\fd6 \, E.
\end{eqnarray}
\normalsize

\begin{figure*}
\scalebox{1}[1]{\includegraphics[angle=0,width=18cm]{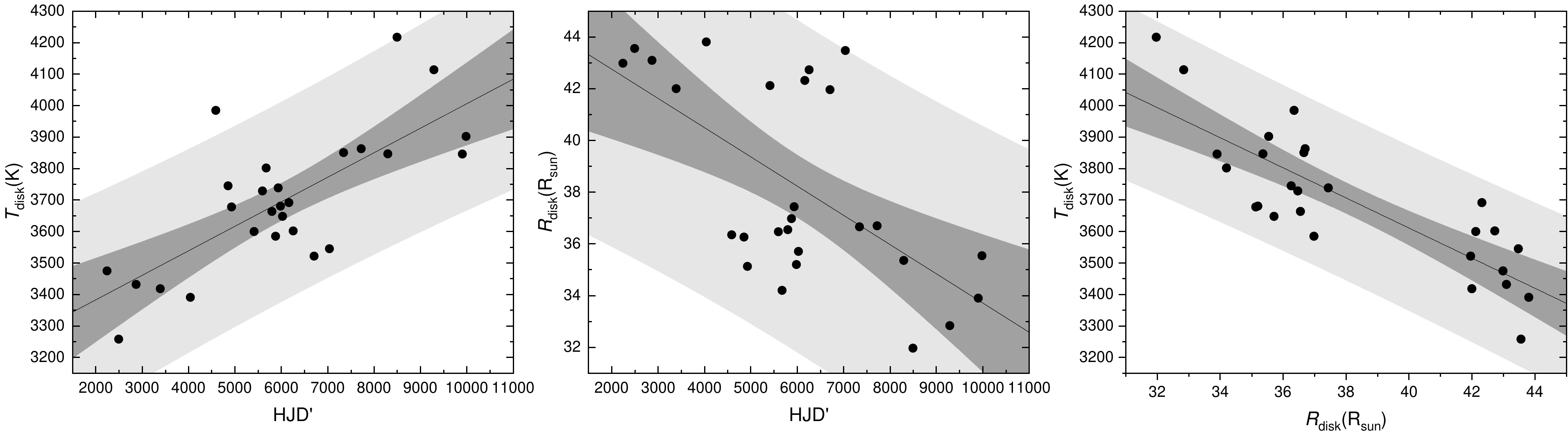}}
\caption{Behaviour of some disk parameters according to our models. Typical errors are 60 K and 0.1\rsun. Best linear fits are shown along with grey light and grey dark regions indicating 95\% prediction and confidence bands, respectively.  Parameters for these fits are given in Table \ref{tab:paramfits}.}
\label{fig:mosaico}
\end{figure*}

\begin{figure}
\scalebox{1}[1]{\includegraphics[angle=0,width=8cm]{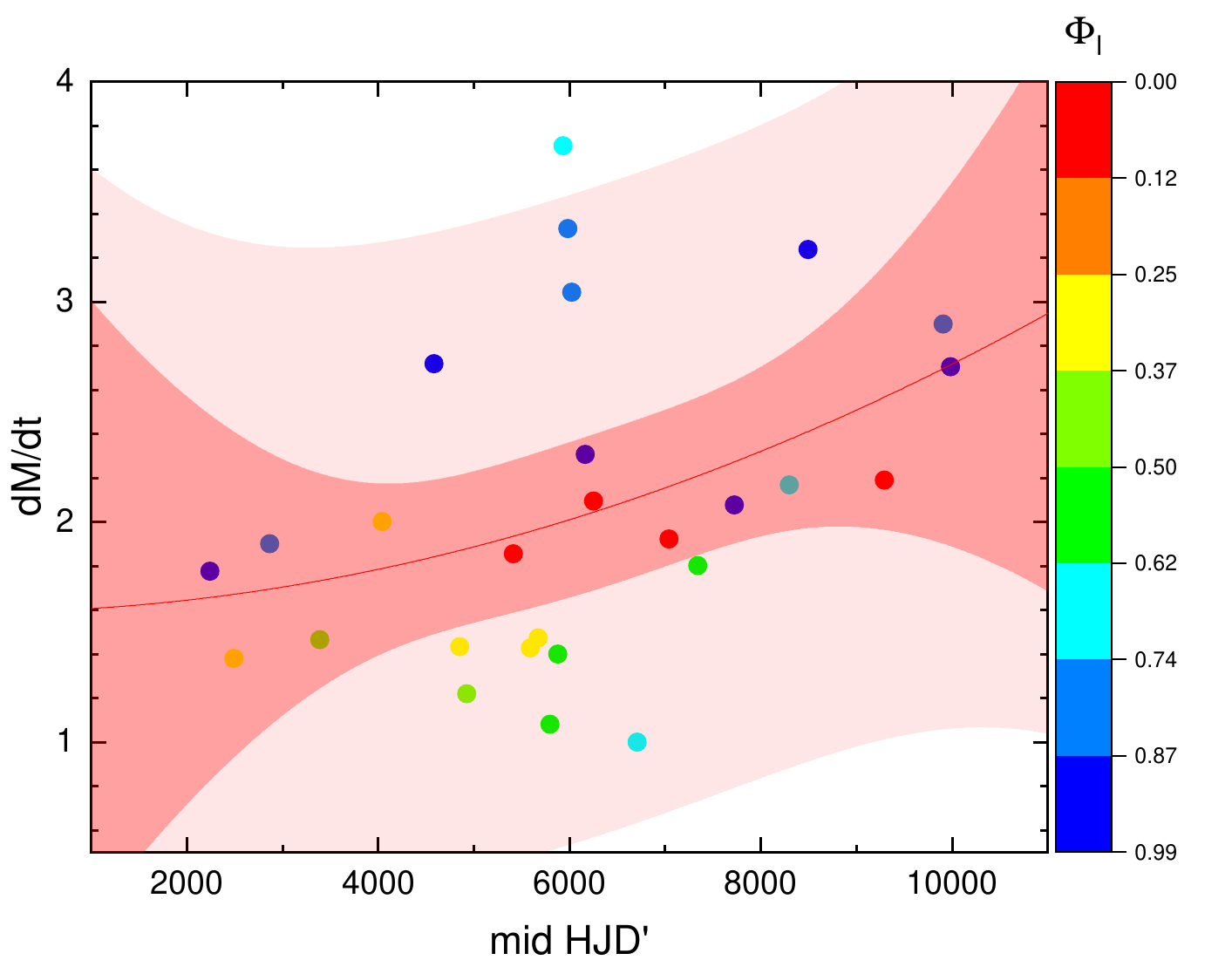}}
\caption{Mass transfer rate versus the mid HJD' given in Table 2 with colors indicating ranges of the  long-term cycle phase. The best quadratic fit,   y =  (1.59055) + (5.948E-6) * x + (1.06792E-8) * x$^2$ is shown, along with grey light and grey colored regions indicating 95\% prediction and confidence bands, respectively.}
\label{fig:long}
\end{figure}

\noindent  Regarding long-term trends, we observe that the disk temperature increases with Heliocentric Julian Date, while the disk radius decreases.  Additionally, we find that the disk tends to be hotter when it is smaller (Fig. \ref{fig:mosaico}).
Throughout the observation period, the mass transfer rate roughly increases (Fig.\, \ref{fig:long}). This observation should be interpreted considering data collected at similar phases of the  long-term cycle. It is specially evident in the data colored with blue in the above figure ($\Phi_{\rm l}$ = 0.87-0.99). Beyond this overarching trend, the parameter dM/dt also mirrors the  long-term cycle, as we will demonstrate next.


\begin{table}
\centering
\caption{Parameters of fits $y = a + b x$  shown in Fig. \ref{fig:mosaico}. }
\label{tab:paramfits}
\begin{tabular}{llrrrr} 
\hline
Plot &Par. & $Value$ & Std &t-value &Prob $>$ |t|  \\
\hline
$T_{d}$-HJD'&	$a$&	3227&	91	&35.3	&7.3E-23\\
$T_{d}$-HJD'&	$b$&0.078&	0.014&	5.43&	1.2E-5\\
$R_{disk}$-HJD'&	$a$&45.02	&1.85&	24.4&6.0E-19\\
$R_{disk}$-HJD'&	$b$&	-0.0011	&3E-4	&-3.9& 6.4E-4\\
$T_{d}$-$R_{disk}$&	$a$&	5524&	247&	22.4&	4.6E-18\\
$T_{d}$-$R_{disk}$&	$b$	&-47.8& 6.4&	-7.4&	8.5E-8\\
\hline
\end{tabular}
\end{table}

\begin{figure}
\scalebox{1}[1]{\includegraphics[angle=0,width=8cm]{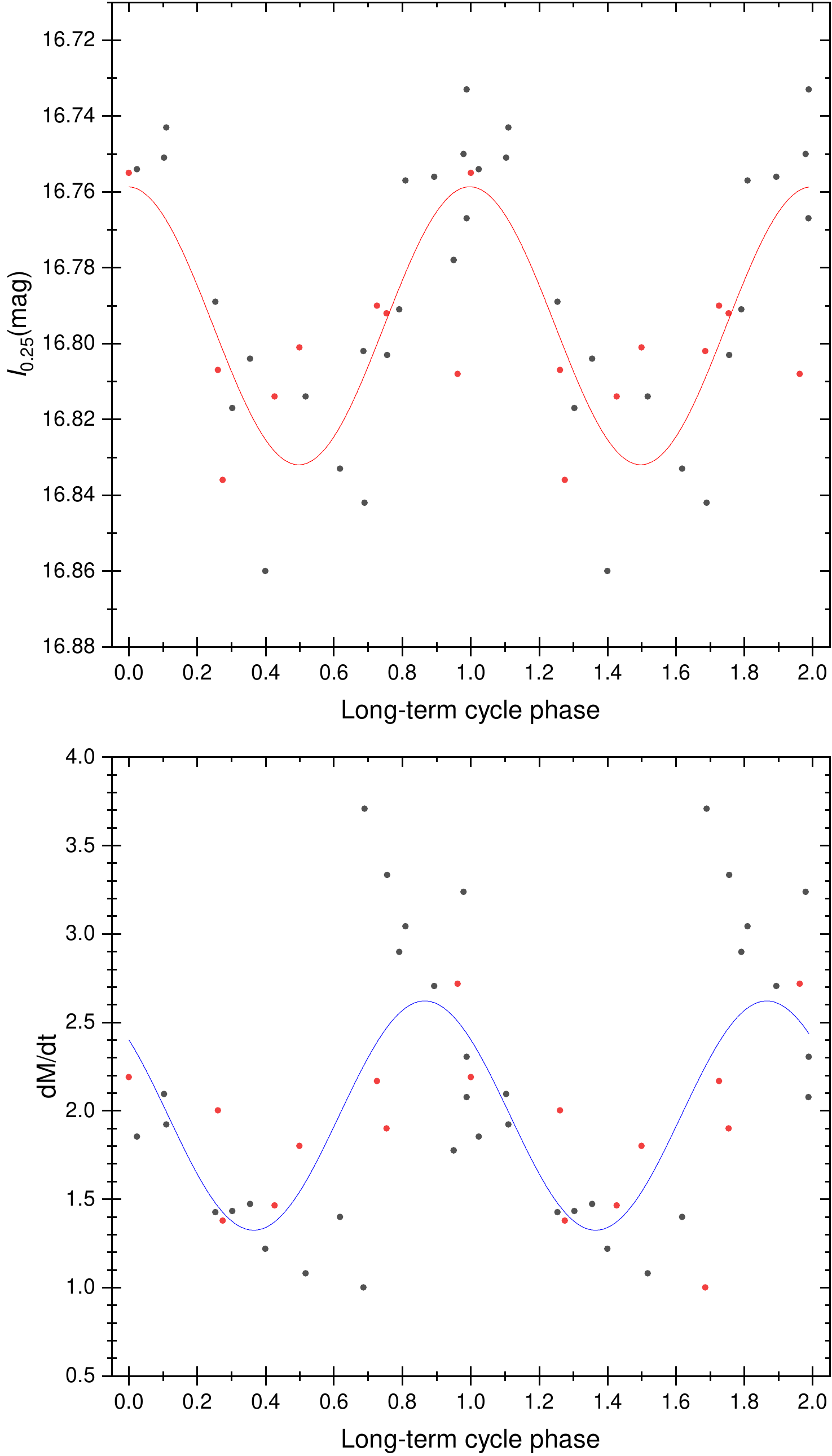}}
\caption{$\dot{M}$ and $I$-band magnitude, measured at orbital phase 0.25, as a function of the  long-term  cycle phase, based on the analysis of the light curve strings of Table \ref{tab:strings}. Red (black) dots correspond to data of datasets whose time span is longer (shorter) than 35\% of the cycle length. Best sinus fits are also shown.}
\label{fig:dotM-mag}
\end{figure}

\begin{figure}
\scalebox{1}[1]{\includegraphics[angle=0,width=8cm]{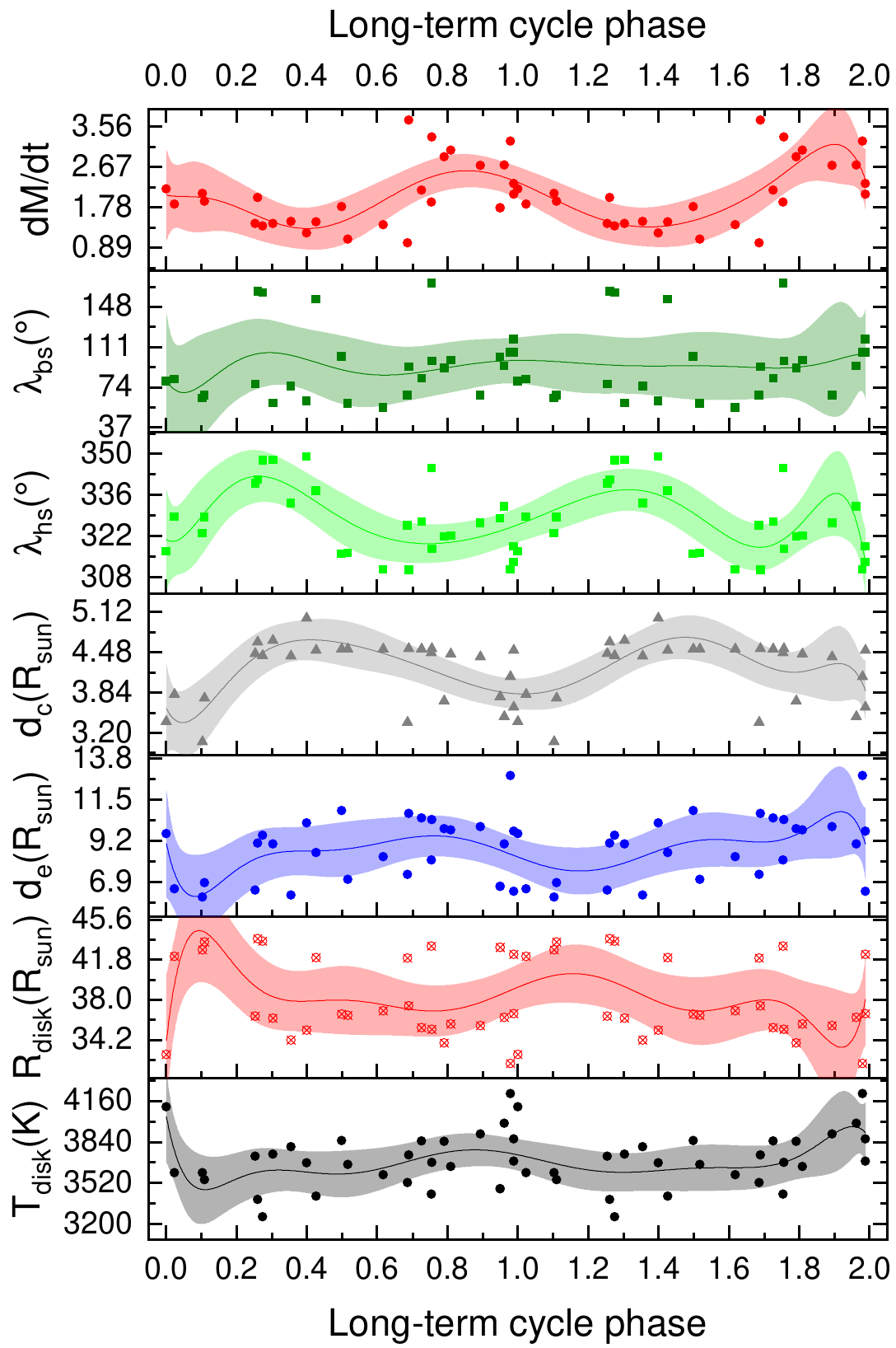}}
\caption{Behaviour of some disk parameters during the  long-term cycle. Polynomial fits of degree 9th are shown to illustrate tendencies.}
\label{fig:mosaicocolor}
\end{figure}

We notice  that the magnitude at orbital phase of 0.25 mirrors the dynamics of the  long-term cycle, similarly to how the mass transfer rate does, as shown in Figure  \ref{fig:dotM-mag}. Both parameters undergo clear changes throughout the  long-term cycle, although with significant scatter. Notably, the peak of the  long-term cycle nearly aligns with the maximum in the mass transfer rate, and both curves exhibit a slight shift where the mass transfer rate ($\dot{M}$) precedes the system's brightness. This suggests a causal relationship: an increase in mass transfer rate leads to an increase in system brightness. The enhanced brightness during the peak of the  long-term cycle could result from reduced obscuration of the gainer star and a hotter disk, as implied by the $d_c$ and $R_{disk}$ diagrams in Fig. \ref{fig:mosaicocolor}. In the same figure we observe that the hot spot position moves closer to the line joining the stars around $\Phi_l$ = 0.3.

\begin{figure}
\scalebox{1}[1]{\includegraphics[angle=0,width=8cm]{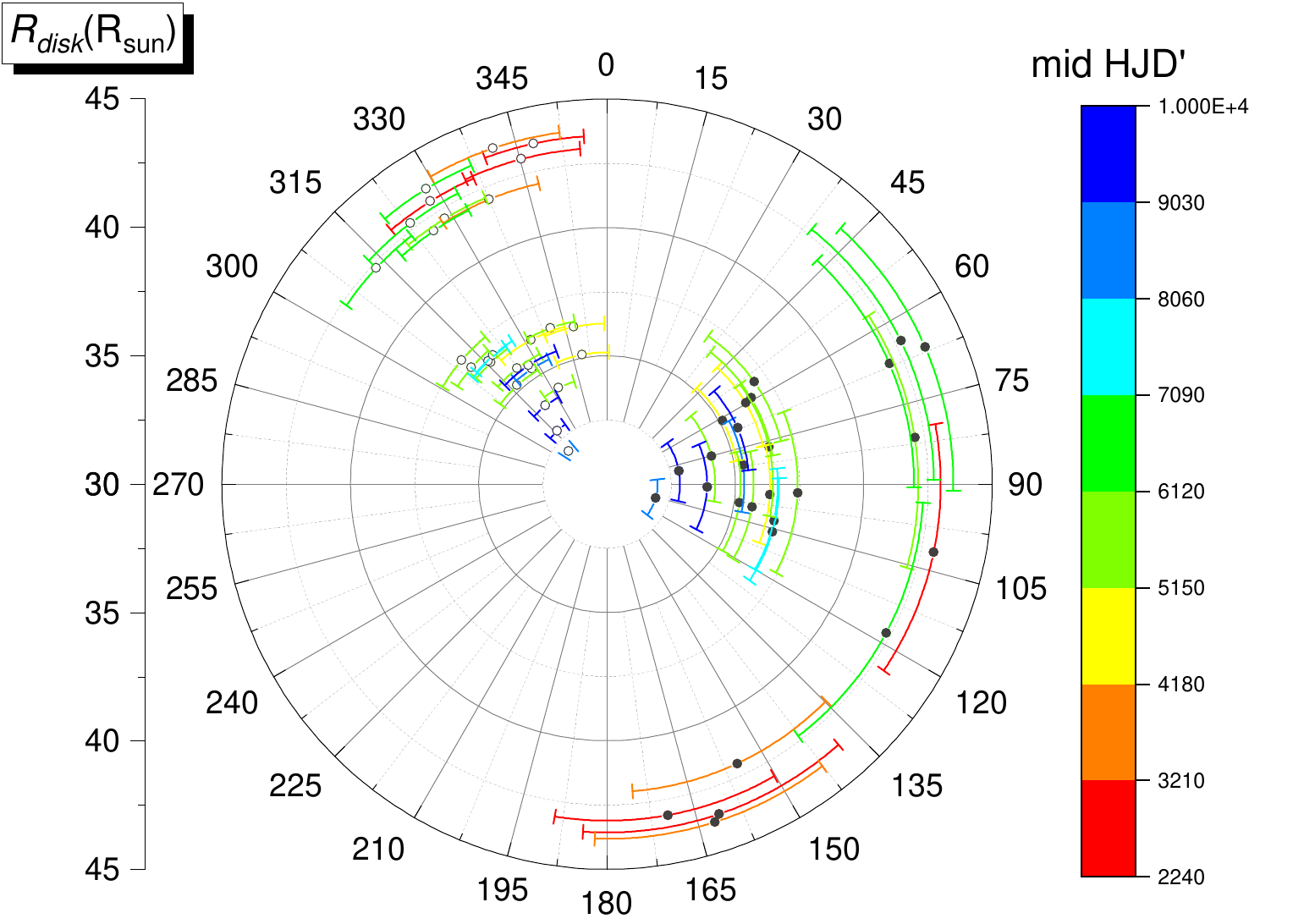}}
\caption{Radial position and angular extension of hot spot (open circles) and bright spot (dots) at different epochs.}
\label{fig:polar}
\end{figure}

Interestingly, the long-term tendency of disk radius can be observed in the polar diagram of Fig.\, \ref{fig:polar}. We see that larger disk are found at the earliest epochs, whereas smaller disk can be found at later epochs. It is at earlier epochs when the bright spot attains larger extension and it is distributed over a wider area in the outer disk.

\section{On the evolutionary stage}

We use the Modules for Experiments in Stellar Astrophysics (\texttt{MESA}), a powerful and widely used suite of open-source computational tools designed for astrophysical research
\citep{2011ApJS..192....3P, 2015ApJS..220...15P}. In particular, when studying binary star evolution, \texttt{MESA} is extremely useful due to its robust and detailed simulations of stellar physics. Crucial for binary star evolution, \texttt{MESA} can simulate interactions between the stars, such as mass transfer, common envelope phases, and tidal effects. These interactions significantly influence the evolution and fate of binary systems. For our purposes, we customize \texttt{MESA} following Rosales et al. (2024), matching the scenario of mass transfer by Roche lobe overflow between two intermediate mass stars. 

We performed several trials with initial stellar masses in a range 3.4$-$5.8 \msun\ for the donor and 1.0$-$3.4 \msun\  for the gainer, both with step of 0.05 \msun\ and orbital periods in a range 1$-$10 days with step 1.0 day, allowing the systems to evolve from its initial condition. We fixed the metallicity to $Z$= 0.006, roughly representative for the  Large Magellanic Cloud \citep{2021A&A...652A.137E}.  We constructed models that were compared with the observed system properties. 
Among these models we selected those that minimized the $\chi^{2}$ parameter, representing mean deviations of the system parameters regarding those of theoretically calculated values. Following Rosales et al. (2024),  
the prescription for mass transfer considered the fraction of mass lost from the vicinity of donor as fast wind as 1E-3, the fraction of mass lost from the vicinity of accretor as fast wind as 1E-4 and no mass lost from a possible circumbinary matter. The first two parameters were allowed to vary from 1E-1 to 1E-12 in steps of 1 dex. These values were chosen as representative of a case with very small or neglectable mass loss,  considering that the more evolved donor should have a wind larger than the B-type dwarf.   In other words, we considered quasi-conservative models. The mixing length parameters for donor and gainer were allowed to vary from 0.3 to 1.9 in steps of 0.1 and the final values chosen were 1.0 both. The parameter for semi convection was fixed to 0.01. The above values guarantee an effective convective mixing of chemical species. In order to explore magnetism in the donor star, we activate rotation in  the \texttt{MESA} code, enabling also thermohaline mixing. On the contrary, in order to keep simple and fast the calculations, we assumed zero rotation for the gainer switching off this mode.

Our studies showed that the present day binary parameters can be reproduced by a binary of initial period 
7 days and initial masses 3.5 and 3.3 \msun\ and current age 200.54 Myr. The evolution of the best model is summarized in Table \ref{table:evolution}, at specific evolutionary stages labeled accordingly, and from the Zero Age Main Sequence to $^{4}$He depletion.  

\begin{figure}
\scalebox{1}[1]{\includegraphics[angle=0,width=8cm]{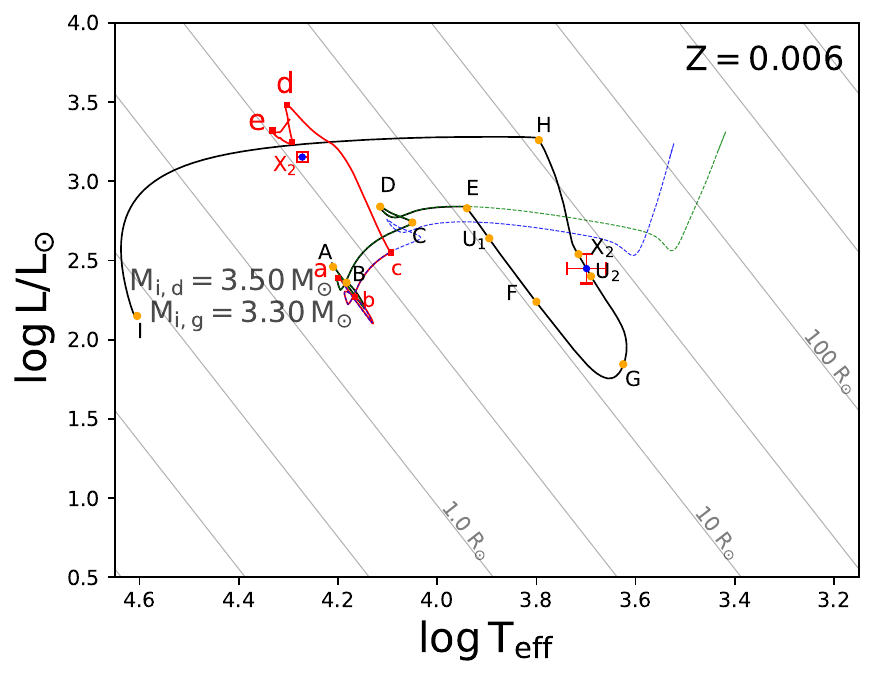}}
\caption{Evolutionary paths of the donor and the gainer for the best model. Model initial and best-fit mass values are shown, along with letters indicating the evolutionary stages described in Table \ref{table:evolution}.}
\label{fig:LT}
\end{figure}

\begin{figure*}
\scalebox{1}[1]{\includegraphics[angle=0,width=8cm]{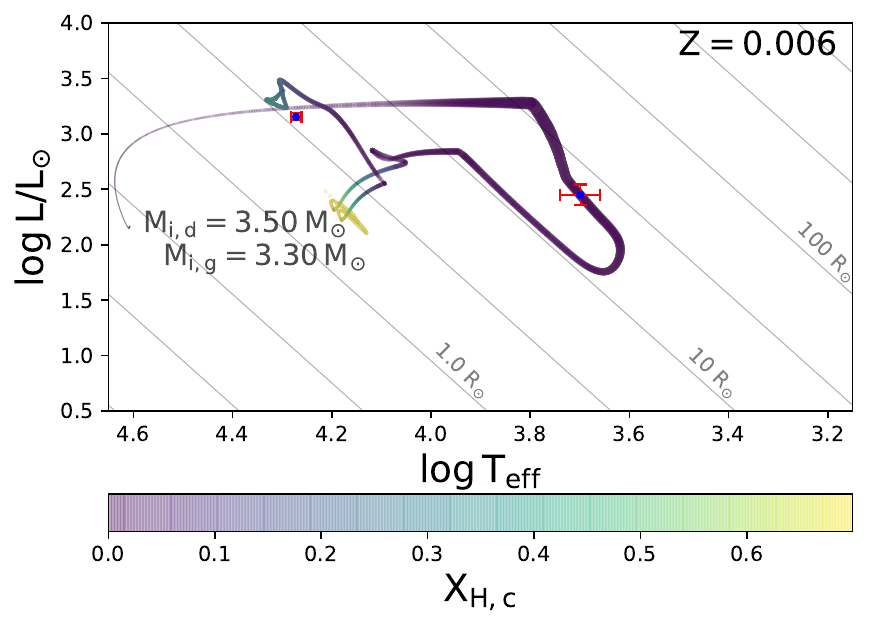}}
\scalebox{1}[1]{\includegraphics[angle=0,width=8cm]{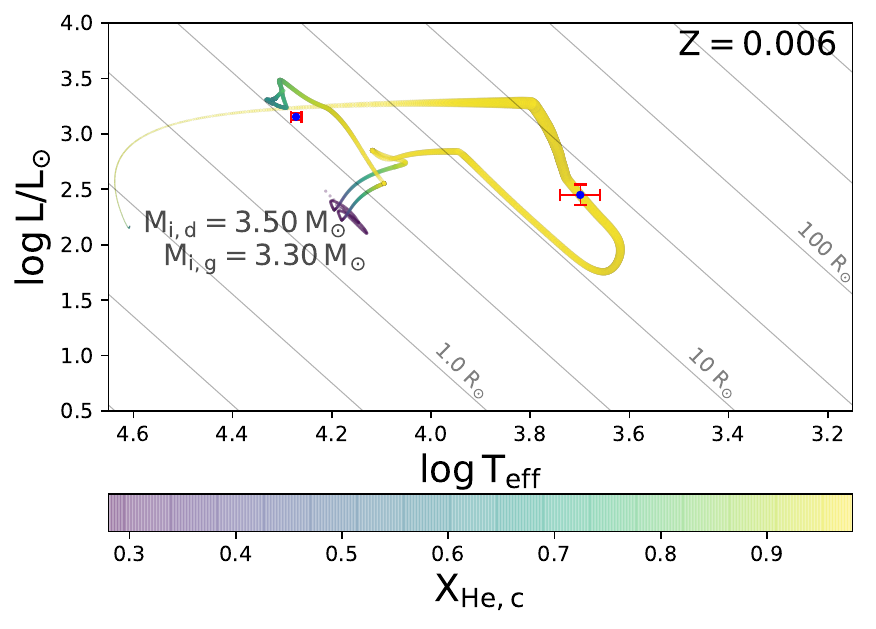}}
\caption{HR diagrams indicating the system evolution along with the hydrogen and helium fractions in the stellar cores.}
\label{fig:LTcore}
\end{figure*}

\begin{figure}
\scalebox{1}[1]{\includegraphics[angle=0,width=8cm]{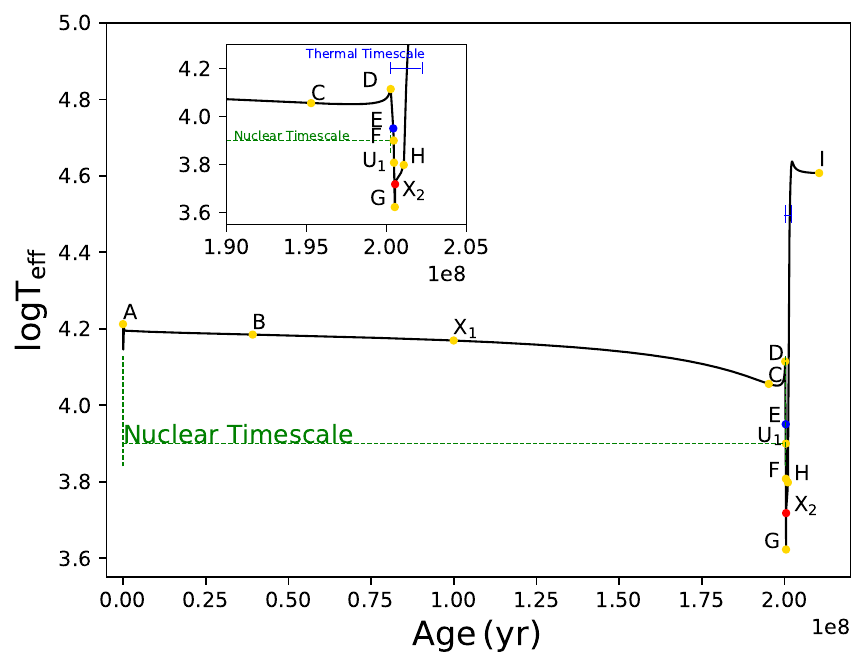}}
\scalebox{1}[1]{\includegraphics[angle=0,width=8cm]{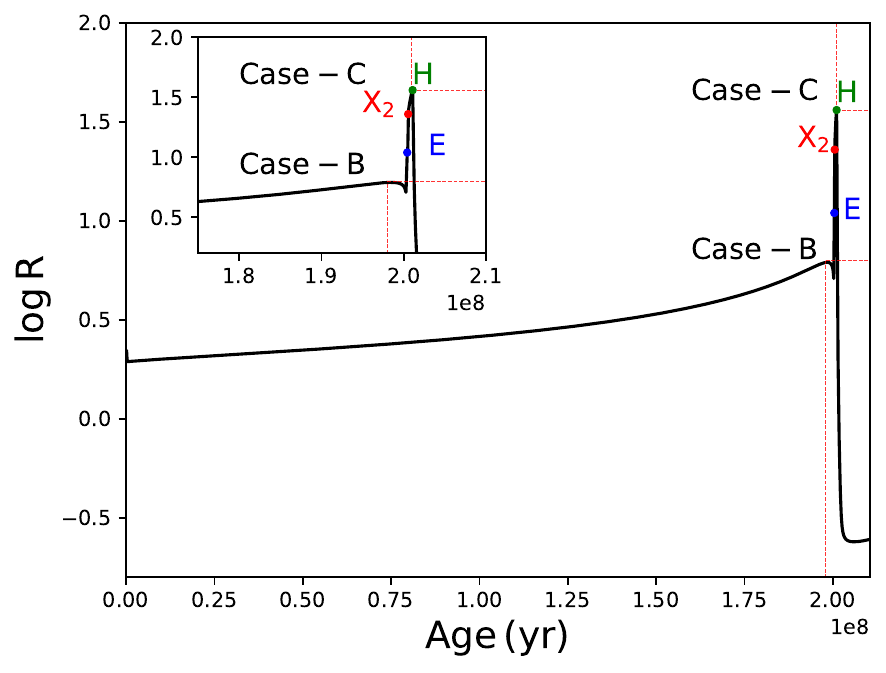}}
\caption{(Left) Evolution of the donor star size with an initial mass of 3.5 \msun\ in the binary system \var, with a companion that has an initial mass of 3.3 \msun, showing the phases in which cases A, B, and C of RLOF can occur, depending on the age of the system. The inversion of the ${}^{1}$H/${}^{4}$He  ratio (X1 stage) is represented by a yellow dot, the mass transfer (E stage) by a blue dot, and the current stage of the system (X2 stage) by a red dot. (Right) Effective temperature $T_{\rm{eff}}$ as function of age for \var. We show that the donor star, after crossing the Hertzsprung gap during the onset the blue loop (Stages C and D), follows an evolutionary track until reaching its current stage (X2 stage) on a thermal timescale. This continues until reaching the final stage of non optically-thick mass transfer (H-stage), stopping shortly before Helium depletion.}
\label{fig:timescales}
\end{figure}

The stellar and system parameters found compare well with those reported in Table \ref{tab:system}; namely, theoretical masses 5.6 and 1.2 \msun\  versus 5.8 and 1.1 \msun\, calculated radii of 23.0 and 3.8 \rsun\ versus 22.4 and 3.6 \rsun\ and orbital period 38\fd16 versus 38\fd16. Regarding the luminosities, the theoretical temperatures are log\,$T$ = 3.72 and 4.29 K, versus log\,$T$ = 3.70 and 4.27 K. We notice a small  discrepancy with the luminosity of the gainer, and this principally produces the mismatch between the observed data and the predicted position (X$_{2}$ stage) in Fig. \ref{fig:LT}. This is observed even at trials with metallicities of Z= 0.02 and 0.003. It can, in principle, be attributed to the presence of the disk, that hides the more massive star and  whose influence in the flux balance has not been included in the \texttt{MESA} calculations.  For the same reason, we observe our simulations as general insight of the possible evolutionary path for the binary, and we do not pursue a better approximation, although it is remarkable that the best model is in fact  found in a mass transfer stage, with $dM/dt$ = 3.5 $\times$ 10$^{-6}$ \msun/yr. The donor is found after hydrogen depletion and with important helium enhancement in its core, while the gainer has rejuvenated acquiring 2.3 \msun\ during the accretion process (Table \ref{table:evolution} and Fig. \ref{fig:LTcore}). The system already passed by the optically thick mass transfer stage and now it is in a milder stage of mass transfer.   During the binary evolution, the donor increased its radius from 2.2 \rsun\  to 23 \rsun\  and the orbital period increased from 7\fd0 to the current value of   38\fd16. The Hertzsprung-Russell (HR) diagrams show the huge decrease of temperature of the donor star, and the increase of luminosity of the gainer, until reaching the current stage. They also show an expected decrease of the hydrogen and the increase of the helium fraction in the core of the gainer and the donor (Fig. \ref{fig:LTcore}). Furthermore, our \texttt{MESA} simulations shows, as expected, how the temperature of the donor evolves first at a nuclear timescale and then, after filling the Roche lobe, at a much shorter thermal time scale when mass transfer occurs. The same for the donor radius, that slightly increases during millions of years to change abruptly,  after starting the mass transfer process (Fig. \ref{fig:timescales}).

\begin{figure}
\scalebox{1}[1]{\includegraphics[angle=0,width=8cm]{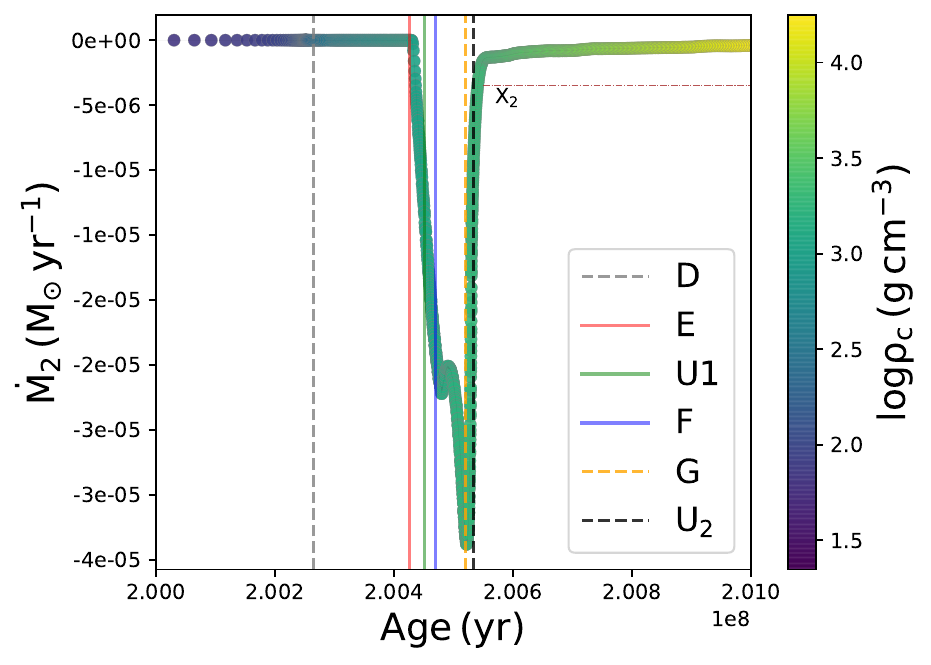}}
\caption{Mass transfer rate  versus the system's age for the best model. Vertical lines indicate the evolutionary stages described in Table \ref{table:evolution}. The density in the core of the donor star is also shown. }
\label{fig:mdot-time}
\end{figure}

We find the system in a mass transfer stage, returning from a huge burst of mass transferred to the gainer at maximum rate of 1 $\times$ 10$^{-4}$ \msun/yr; at present time the rate of mass transfer has slowed to  $dM/dt$= 3.46  $\times$ 10$^{-6}$   \msun/yr. The overall process, from stages E to H,  will last about 655\,000 years (Fig. \ref{fig:mdot-time}). 

\begin{figure}
\begin{center}
\scalebox{1}[1]{\includegraphics[angle=0,width=8cm]{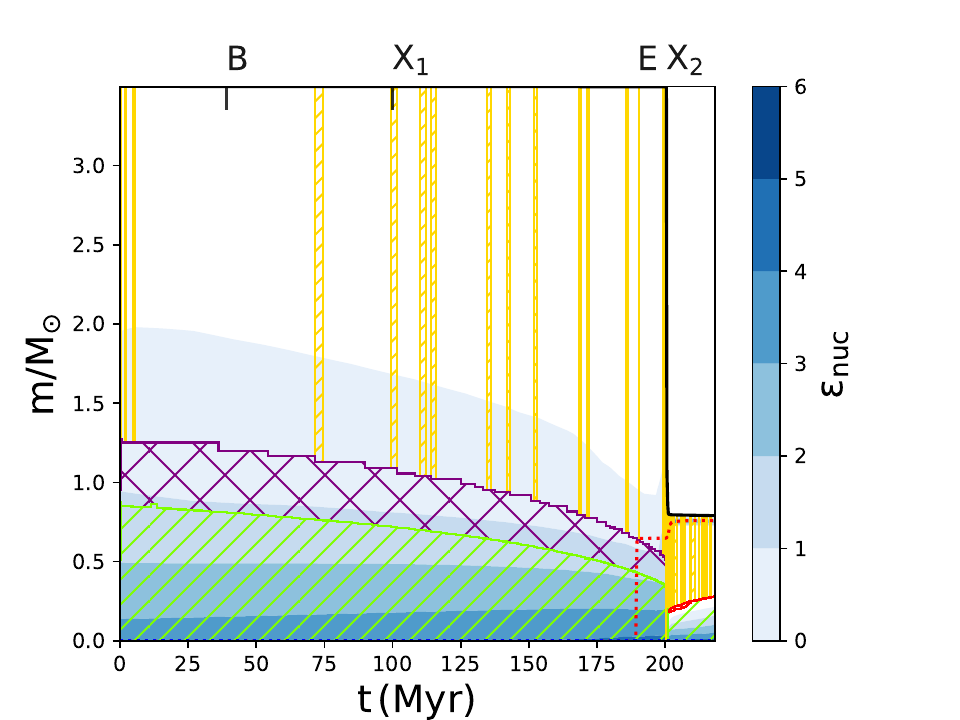}}
\scalebox{1}[1]{\includegraphics[angle=0,width=8cm]{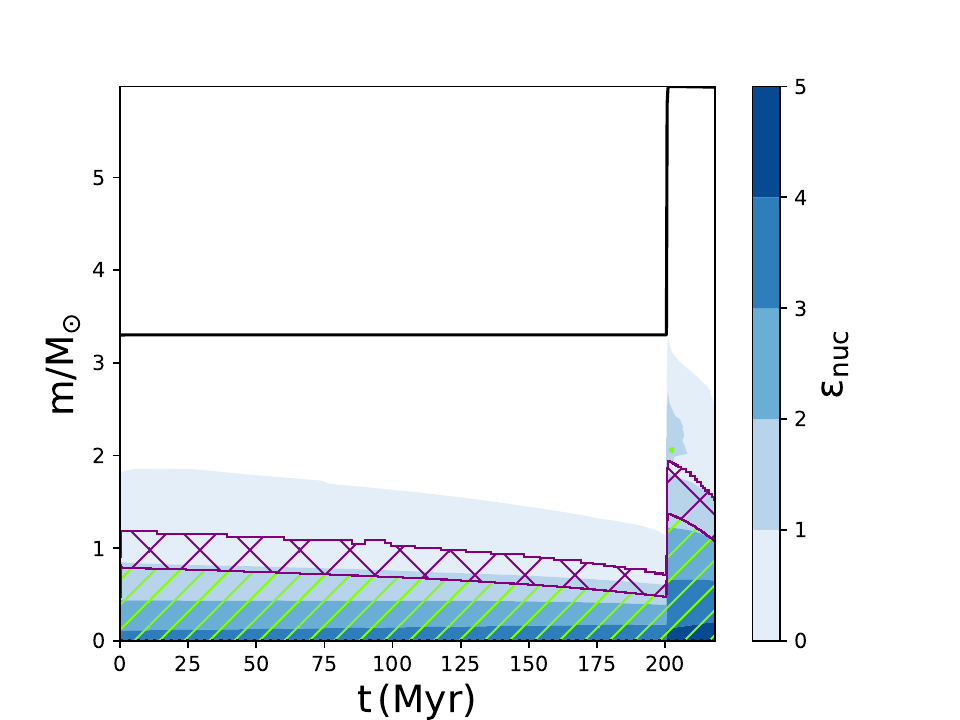}}
\caption{Kippenhahn diagrams showing the internal structure of
the donor (up) and gainer (down) stars. Diagrams with initial masses 3.5 and 3.3 \msun\ are shown. 
The evolutionary calculations were stopped when the donor star reached core helium depletion X(He$_c$) $<$ 0.2.  The x-axis gives the
age after ignition of hydrogen in units of Myr. The different layers are characterized by their values of M/\msun. The convection
mixing region is represented by hatched green, semi convection mixing region 
in red and the overshooting mixing region in crosshatched purple. The red
dots represent the He core mass while the thermohaline mixing region is represented in hatched
yellow. The solid black lines show the surfaces of the stars.}
\label{fig:Kipp}
\end{center}
\end{figure}

We further analyze the internal structure of both the donor and the gainer stars using Kippenhahn diagrams (Fig. \ref{fig:Kipp}). For the case of the donor, which starts from the ZAMS (Zero Age Main Sequence) until helium depletion X(He$_{\rm c}$) $<$ 0.2, we were able to identify that the convective zone of the donor
star is below 0.5 \msun\ while mostly all nuclear production occurs below 1 \msun. Additionally, both zones of convection and nuclear production gradually
decrease as the system evolves and approaches mass transfer as expected (E stage). Zones of thermohaline instability activation intermittently appear at different stages of evolution. In addition, the overshooting zone is present only until mass transfer occurs and then disappears once mass transfer begin. 

For the gainer star, we identified a convective zone from its center to 0.5 \msun\ with a nuclear production rate zone that remains largely unchanged
for most of its lifetime from the ZAMS (Zero Age Main Sequence) to moments prior to mass transfer (E-stage). In addition, we identified an overshooting zone from 0.7 to 0.9 \msun. However, after mass transfer, its convective core immediately grew twice the original size, in terms of total mass. Following this, the overshooting zone appears between 1.5 and 2.0 \msun. Additionally, the nuclear production  zone increased significantly, occurring within a central mass of 3 \msun\ after mass transfer. 

\begin{figure*}
\scalebox{1}[1]{\includegraphics[angle=0,width=6cm]{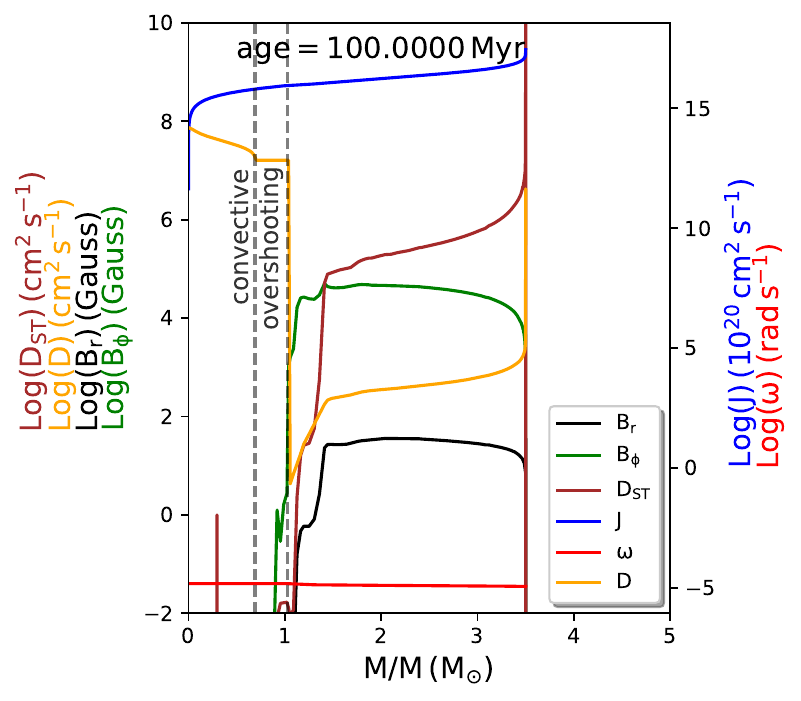}}
\scalebox{1}[1]{\includegraphics[angle=0,width=6cm]{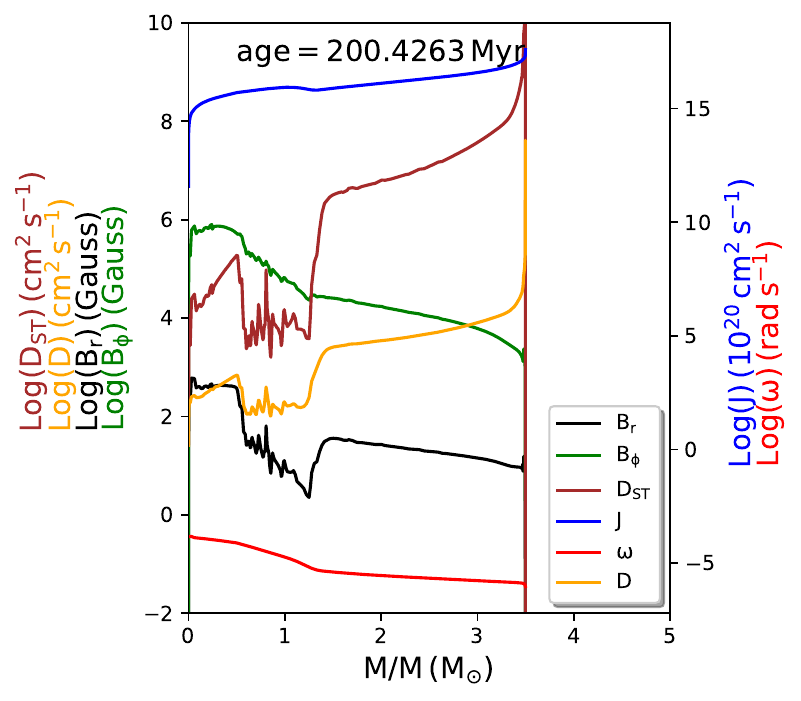}}
\scalebox{1}[1]{\includegraphics[angle=0,width=6cm]{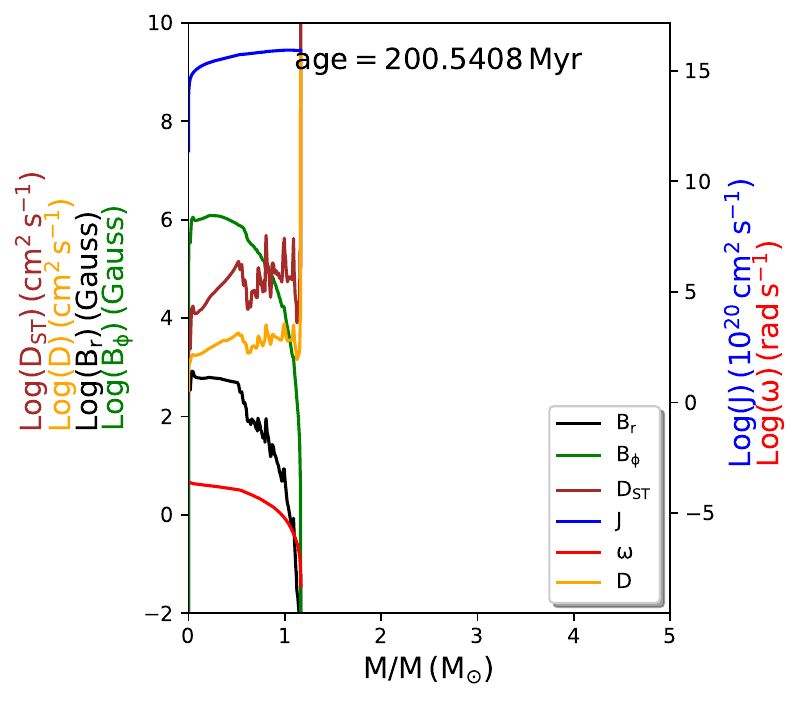}}
\caption{Profiles of magnetic fields generated by the Tayler-Spruit dynamo in the poloidal B$_r$ and toroidal B$_t$ components for the donor star at different evolutionary stages. The profiles represent the stages of inversion of the $^{1}$H/$^{4}$He ratio (X$_{1}$ stage, left panel), the start of mass transfer (E-stage, middle panel) and  the current stage (X2, right panel). The Eulerian diffusion coefficient for mixing is represented in orange color, the Spruit-Tayler (ST) diffusion coefficient by the brown line, the poloidal magnetic field as a black line, the toroidal magnetic field in green, whereas the angular momentum is represented in blue and the angular velocity  in red.  }
\label{fig:fields}
\end{figure*}

We calculated the magnetic field structure due to the Taylor-Spruit (ST) dynamo, in particular the toroidal and poloidal components, that resulted of the order of few Gauss in the stellar subsurface. On the other hand, the Eulerian diffusion for mixing and the Spruit-Taylor diffusion coefficient and the angular velocity were also calculated and are shown in Fig. \ref{fig:fields}. The magnetic fields obtained with the ST prescription, of the order of few Gauss, are too weak to be detected observationally with the  current usual instrumentation.

\begin{table*}[h!]
\caption{Evolutionary stages of \var.}
\normalsize
\begin{center}
\resizebox{0.95\textwidth}{4.0cm}{
\begin{tabular}{llrrrrcrl}
	\hline
	\noalign
	{\smallskip}
	\textrm{}       & \textrm{Stage}    & \textrm{Age (Myr)} & \textrm{M (M$_{\odot}$)} & \textrm{R (R$_{\odot}$)}  & \textrm{P$_{\rm{o}}$ (d)}  & \textrm{$\log{T}$ (K)} & \textrm{$\mathrm{\dot{M}}$\,(M$_{\odot}$ yr$^{-1}$)}   & \textrm{Ev. process}             \\
	\hline
\noalign{\smallskip}
\textrm{Donor}  & \textrm{A}        & \textrm{  0.0000}  & \textrm{3.5000}         & \textrm{2.1860}     & \textrm{7.0000}  & \textrm{4.2122}& \textrm{-1.807E-12}        & \textrm{Zero Age Main Sequence (ZAMS)}\\
\textrm{}       & \textrm{B}        & \textrm{ 39.2148}  & \textrm{3.4999}         & \textrm{2.1619}     & \textrm{7.0005}  & \textrm{4.1848}& \textrm{-3.953e-12}        & \textrm{Terminal Age Main Sequences (TAMS)}\\
\textrm{}       & \textrm{X$_{1}$}  & \textrm{100.0000}  & \textrm{3.4995}         & \textrm{2.6076}     & \textrm{7.0014}  & \textrm{4.1694}& \textrm{-6.338E-12}        & \textrm{Inversion $\rm{{}^{1}H/{}^{4}He}$ donor}\\
\textrm{}       & \textrm{C}        & \textrm{195.2898}  & \textrm{3.4983}         & \textrm{5.9322}     & \textrm{7.0050}  & \textrm{4.0558}& \textrm{-7.298E-11}        & \textrm{Depletion of central hydrogen $\rm{{}^{1}H}$}\\ 
\textrm{}       & \textrm{D}        & \textrm{200.2644}  & \textrm{3.4979}         & \textrm{5.1146}     & \textrm{7.0059}  & \textrm{4.1145}& \textrm{-3.715E-11}        & \textrm{Size increase beyond the RL due to depletion of 1 H}\\
\textrm{}       & \textrm{E}        & \textrm{200.4263}  & \textrm{3.4979}         & \textrm{11.0219}    & \textrm{7.0060}  & \textrm{3.9503}& \textrm{-7.255E-10}        & \textrm{Initiation of mass transfer}\\
\textrm{}       & \textrm{U$_{1}$}  & \textrm{200.4512}  & \textrm{3.3590}         & \textrm{11.2709}    & \textrm{6.9961}  & \textrm{3.8998}& \textrm{-1.351E-05}        & \textrm{Mass inversion ($\rm{M_{1}=M_{2}}$)}\\
\textrm{}       & \textrm{F}        & \textrm{200.4700}  & \textrm{2.9999}         & \textrm{10.9856}    & \textrm{7.2961}  & \textrm{3.8076}& \textrm{-2.406E-05}        & \textrm{Minimun value Roche Lobe}\\
\textrm{}       & \textrm{G}        & \textrm{200.5200}  & \textrm{1.5973}         & \textrm{16.0870}    & \textrm{18.8282} & \textrm{3.6227}& \textrm{-3.900E-05}        & \textrm{Maximum mass transfer}\\
\textrm{}       & \textrm{U$_{2}$}  & \textrm{200.5333}  & \textrm{1.2091}         & \textrm{21.9994}    & \textrm{34.9631} & \textrm{3.6890}& \textrm{-1.004E-05}        & \textrm{End of optically thick mass transfer}\\
\textrm{}       & \textrm{X$_{2}$}  & \textrm{200.5408}  & \textrm{1.1650}         & \textrm{22.9686}    & \textrm{38.1664} & \textrm{3.7179}& \textrm{-3.458E-06}        & \textrm{Current stage}\\
\textrm{}       & \textrm{H}        & \textrm{201.0810}  & \textrm{0.80231}        & \textrm{36.5068}    & \textrm{94.9661} & \textrm{3.7982}& \textrm{-7.685E-08}        & \textrm{end mass transfer}\\
\textrm{}       & \textrm{I}        & \textrm{210.5306}  & \textrm{0.7926}         & \textrm{0.24643}    & \textrm{95.3674} & \textrm{4.6073}& \textrm{-3.059E-10}        & \textrm{$\mathrm{{}^{4}He}$ depletion}\\
\hline
\textrm{Gainer} & \textrm{}         & \textrm{}  		 & \textrm{}    	 	   & \textrm{}           & \textrm{}  		& \textrm{}      & \textrm{}				  & \textrm{} \\
\textrm{}       & \textrm{a}        & \textrm{  0.0000}  & \textrm{3.3000}   	   & \textrm{2.1194}     & \textrm{7.0000}  & \textrm{4.1981}& \textrm{-1.049E-12}        & \textrm{Zero Age Main Sequence (ZAMS)}\\
\textrm{}       & \textrm{b}        & \textrm{ 39.2148}  & \textrm{3.2999}   	   & \textrm{2.0593}     & \textrm{7.0005}  & \textrm{4.1706}& \textrm{-2.449E-12}        & \textrm{Terminal Age Main Sequences (TAMS)}\\
\textrm{}       & \textrm{c}        & \textrm{200.4263}  & \textrm{3.2992}         & \textrm{4.0710}     & \textrm{7.0060}  & \textrm{4.0943}& \textrm{-8.432E-12}        & \textrm{Initiation of mass accretion}\\
\textrm{}       & \textrm{U$_{1}$}  & \textrm{200.4512}  & \textrm{3.4379}         & \textrm{4.664}      & \textrm{6.9961}  & \textrm{4.1410}& \textrm{ 1.349E-05}        & \textrm{Mass inversion ($\rm{M_{1}=M_{2}}$)}\\
\textrm{}       & \textrm{d}        & \textrm{200.5200}  & \textrm{5.1976}         & \textrm{4.6279}     & \textrm{18.8282} & \textrm{4.2924}& \textrm{ 3.895E-05}        & \textrm{Maximum mass accretion and relocation to HR diagram}\\
\textrm{}       & \textrm{U$_{2}$}  & \textrm{200.5333}  & \textrm{5.5854}         & \textrm{4.0368}     & \textrm{34.9631} & \textrm{4.3001}& \textrm{ 1.003E-05}        & \textrm{End of optically thick mass accretion}\\
\textrm{}       & \textrm{X$_{2}$}    & \textrm{200.5408}  & \textrm{5.6294}         & \textrm{3.7647}     & \textrm{38.1664} & \textrm{4.2946}& \textrm{ 3.451E-06}        & \textrm{Current stage}\\
\textrm{}       & \textrm{e}        & \textrm{201.0810}  & \textrm{5.9815}         & \textrm{3.3385}     & \textrm{94.9661} & \textrm{4.3269}& \textrm{ 3.408E-08}        & \textrm{End mass accretion}\\
\hline
\end{tabular}}
\end{center}
\label{table:evolution}
\vspace{0.05cm}
Note: Evolutionary stages of \var, from Zero Age Main Sequence (ZAMS) to ${}^{4}$He depletion of the donor star, according to the best model. Detailed descriptions of key features are provided, along with corresponding ages measured in Mega years (Myr). 
\end{table*}

\section{Discussion}

Before we proceed with the interpretation of our results, it's important to be aware of the limitations of our model. A key simplification is our focus exclusively on circumstellar material within the disk, thereby overlooking any light contributions originating from regions above or below the disk plane. Additionally,  we have assumed 
the gainer is surrounded by an accretion disk, disregarding other potential light sources such as jets, winds, and outflows. Nonetheless, the model allows for the possibility of accounting for variations in disk emissivity that are dependent on azimuthal position through the inclusion of two disk spots. Despite these constraints, and drawing on prior research on algols with disks, our model likely captures the principal light sources for the continuum light. This is evidenced by the close correspondence between the model light curve and the orbital and long-term light curves over the span of  21.7 years of observation (the time covered by the $I$-band OGLE time series).  In addition, we run evolutionary models and find the best one reproducing the present state of the system. Although the match is not perfect, we can certainly infer that the system is product of the evolution of a stellar pair that experienced  strong mass transfer in the past, or is still in a semidetached stage with Roche lobe overflowing. 

The fractional radius of the gainer is $R_1/a$ = 0.04. This value reveals that there is enough room around the gainer for the formation of an accretion disk, since for the system mass ratio, the accreting star is smaller than the circularization radius, where a particle released at the inner Lagrangian point arrives to conserve its angular momentum \citep{1975ApJ...198..383L}. The low value of the fractional radius is due to the large orbital separation and long orbital period of this system and it is consistent with the existence of an accretion disk inferred from the light curve model. It is comparable to the value obtained for the long period system V495\,Cen \citep{2018MNRAS.476.3039R, 2021AJ....162...66R}.  Interestingly, in \var, the disk vertical height near the star is larger than twice the gainer radius, provoking a significant occultation of the star.
 
During the 30.85 year time baseline the system shows systematic changes of disk properties, in particular its vertical height at the outer edge.
In hydrostatic equilibrium, the vertical height $H$ of an accretion disk at radius $R_d$ is:

\begin{equation}
\frac{H}{R_d} \approx \frac{c_s}{v_k} = c_s \sqrt{\frac{R_d}{GM_1}},
\end{equation}

\noindent
where $v_k$ is the Keplerian velocity and $c_s$ the sound speed, that for an isothermal perfect gas can be approximated as:\\
 \begin{equation} c_s \approx 10 \sqrt{\frac{T}{10^4\,K}} \, \rm{km\,s^{-1}},
\end{equation}

\noindent
\citep[Eqs. 3.35 and 3.32 in][]{2010eea..book.....K}.
Using the mean parameters at the outer and inner disk we get ($v_k,c_s$) = (164, 6.1) and (552, 13.7) in km\,s$^{-1}$, respectively, yielding $H/R$ = 0.04 and 0.02.  Considering the averages $d_e/R_d$ = 0.23 and $d_c/R_1$ = 1.15 and that the vertical thickness is twice $H$, we conclude that the disk vertical height is larger than expected for hydrodynamical equilibrium, at the inner and outer boundaries, suggesting  that turbulent motions dominate the disk vertical structure. In this context, it may be worth mentioning that in a detailed model of $\beta$ Lyrae, the disk height had to be multiplied  by a factor about 4 times its expected equilibrium value, possibly reflecting non-negligible hydrodynamic flows within the disk \citep{2021A&A...645A..51B}. These turbulent motions could be produced by the continuos injection of mass from the inner Lagrangian point, through a gas stream, into the accretion disk.  Incidentally, the variable thick disk explains why the primary and secondary eclipses relative depth reverses, since part of the gainer is hidden by the disk that masks its luminosity to a variable degree. In addition, the variable projected disk area into the visual line of sight contributes to occult a fraction of the donor while it is eclipsed. The low contribution of the gainer to the total flux is reproduced by our models as shown in \href{https://doi.org/10.5281/zenodo.14192345}{Figs. A1-A3}. 

In principle, the disk can be stable until the last non-intersecting orbit defined by the tidal radius \citep[][Eq. 2.61]{1977ApJ...216..822P, 1995CAS....28.....W}:

\begin{eqnarray}
\frac{R_{\rm{t}}}{a_{\rm{orb}}}= \frac{0.6}{1 + q},
\end{eqnarray}

\noindent we get $R_t/a_{orb}$ = 0.50, or $R_t$ = 45.4 $R_{\odot}$.  We observe that  during all the observing epochs the disk outer radius keeps inside the volume defined by the tidal radius. 

The value of the mass transfer rate found 3.5 $\times$ 10$^{-6}$ \msun/yr, must be consensual with the constancy of the orbital period. For a binary system ongoing conservative mass transfer, we should expect a change of the orbital period of \citep{1963ApJ...138..471H}:

\begin{equation}
\frac{\dot{P_{\rm o}}}{P_{\rm o}} = 3\dot{M_2}(\frac{1}{M_2}-\frac{1}{M_1}) = 7.6 \times 10^{-6} \rm{yr^{-1}}
\end{equation}

\noindent This means a change of 2.9 $\times$ 10$^{-4}$ days per year, or 8.9 $\times$ 10$^{-3}$ days during the whole observing window. This is not observed, therefore the actual mass transfer rate is smaller than 3.5 $\times$ 10$^{-6}$ \msun/yr or systemic mass loss is important for the angular momentum balance. 

As previously said, a magnetic dynamo in the donor has been invoked as the origin of the  long-term  cycle.  The idea behind this scenario is that the dynamo produces cyclic changes in the equatorial radius of the donor star, enhancing quasi-periodically the amount of mass transferred onto the gainer.   This should produce effects in the accretion disk, that is exposed to cycles of enhanced mass transfer. These effects could be changes in the disk radius, height or even disk temperature. An insight favoring the dynamo hypothesis is the discovery that the donor dynamo number 
increases during epochs of high mass transfer in DPVs \citep{2018PASP..130i4203M, 2019BAAA...61..107S}. In addition, the calculated long period for \var, according to the formula given by \citet{2017A&A...602A.109S}, is 820 days, close to the values obtained from the periodograms and the WWZ analysis. However, our \texttt{MESA} analysis shows that the  magnetic fields, at least those calculated from the Spruit-Taylor prescription, are usually less than a thousand of Gauss, i.e. too low to produce observable effects, as we will show now. 

It has been shown that fluctuations in the radius of a star can be attributed to the gravitational interaction between the orbit and changes in the configuration of a magnetically active star within the binary system. As the star progresses through its activity cycle, variations in the distribution of angular momentum lead to the star's changing shape. Variations in the overfill factor of the Roche lobe should produce changes in mass transfer rate. This was proposed as a possible origin for the DPV  long-term cycles by \citet{2017A&A...602A.109S}.  The change in the stellar radius  $\Delta R$ = $R_{new}$ - $R$, derived from the prescription of \citet[][eqs. 7 and 23]{1992ApJ...385..621A} is given by:

\begin{equation}
\frac{\Delta R}{R} = \frac{2}{3} \frac{R^4}{GM^2} B^2
\end{equation}

\noindent where $B$ is the subsurface magnetic field strength. On the other hand, the expected mass transfer rate is given by  \citep{1988A&A...202...93R}:

\begin{equation}
\dot{M} = \dot{M_0} e^{{\frac{\Delta R}{H_2}}}
\end{equation}

\noindent where $\dot{M_0}$ is the mass transfer rate when $B$ = 0, and it can be assumed as constant, whereas  $H_2$ is the pressure scale height  in the Roche potential given by:

\begin{equation}
H_2 = \frac{k T \mu}{m_p g}  
\end{equation}

\noindent where $k$  is the Boltzmann constant, $T$ the donor temperature, $m_p$ the proton mass, $g$ the local acceleration of gravity, and $\mu$ is the mean molecular weight, assumed 1.23 for a star in the LMC with $Z$ $\approx$ 0.01. In addition, from equations 8 and 9 we get:

\begin{equation}
\dot{M} = \dot{M_0}  e^{ {\frac{2}{3} \frac{R^5}{H_2 GM^2}}  B^2}
\end{equation}

\noindent 
Considering values of $R$ = 22.42 \rsun, $M$ = 1.1 \msun\ and $T$ = 5000 K we get $H_2$ = 56\,284\,560 m, i.e. $R/H_2$ = 277.4.
If we furthermore consider a subsurface magnetic field strength of 10kG, we get a fractional 
radius change $\Delta R/R$ of order of 7.3E-6 and a  change in mass transfer rate by a factor exp(2.06E-3),  i.e.
negligible. Therefore, we need larger magnetic field strengths than those produced by the Spruit-Taylor mechanism in the donor subsurface to produce appreciable changes in mass transfer rates and in particular, to reproduce the broad range of values shown in Table \ref{tab:strings}. 
New investigations of the 3D internal magnetic structure and its dependence on rapid rotation and its temporal evolution should be important to shed light on the importance of stellar magnetism for the DPV  long-term cycles, possibly following the techniques and methodologies developed by  \citet{2023A&A...678A...9N}.

\section{Conclusion}

We have investigated the intriguing light curve of the Double Periodic Variable \var, spanning 30.85 years. Our model successfully reproduces the overall $I$-band light curve, capturing both the orbital variability with a periodicity of 38\fd16 and the DPV cycle, which has a duration of approximately 780 days and exhibits a full amplitude of 0\fm076 at the $I$-band. This model includes the light contribution from an accretion disk, characterized by an average radius of 38 \rsun, and two hot shock regions located at the outer disk edge. Additionally, for the first time, we have measured fundamental stellar parameters for this system. Our findings include stellar components with masses of 5.8 $\pm$  0.3 and 1.1 $\pm$  0.1 \msun, radii of 3.6 $\pm$  0.1 and 22.4 $\pm$  0.8 \rsun, and temperatures of 18701 $\pm$ 208 K and 5000  $\pm$ 200 K, respectively. These stars are separated by a distance of 90.7 $\pm$  1.8 \rsun, and exhibit surface gravities (log\,g) of  4.08 $\pm$ 0.04 and 1.78 $\pm$ 0.03. 

Our findings trace the evolutionary history of the system, revealing that it is currently experiencing a burst of mass transfer as the donor fills its Roche lobe. We also observe significant changes in the disk's vertical height, suggesting that turbulent motions contribute to the disk's vertical extension. The system comprises a B-type dwarf, approximately spectral type B3, accreting matter from a cooler, K5-type giant star, according to the spectral classification scheme given by \citet{1988BAICz..39..329H}. Our \texttt{MESA} simulations, employing the Spruit-Taylor prescription for magnetism, indicate only weak magnetic fields on the cooler star's subsurface, insufficient to account for the observed changes in mass transfer rate.

In terms of the broader context of Double Periodic Variables, our work aligns with previous studies such as those of OGLE-LMC-DPV-097 and OGLE-BLG-ECL-157529 \citep{Garces2018, 2021A&A...653A..89M}, which have also noted long-term changes in accretion disk properties that influence the orbital light curve's shape. Specifically, the maximum brightness during the  long-term cycle in both OGLE-BLG-ECL-157529 and \var\ can be attributed to less obscuration of the gainer by a thinner, hotter disk, thereby increasing the visible surface of the gainer. 
For \var, the most significant finding is that the  long-term cycle could be driven by changes in mass transfer rate, as illustrated in Fig. \ref{fig:dotM-mag}, and that the peak of the  long-term cycle occurs when the disk is hotter and thinner. 

The DPV phenomenon is both fascinating and inherently complex. Future investigations involving high spectral resolution data and decade-long time series are poised to illuminate this phenomenon further. Moreover, new theoretical simulations that leverage increased computational power will enhance our understanding of these binary systems.

\section{Data availability}

Figs. A1-A3 are available at:\\
 https://doi.org/10.5281/zenodo.14192345
 
\begin{acknowledgements}
 We thanks an anonymous referee for the useful comments and suggestions on the first version of this manuscript. We acknowledge support by the ANID BASAL project Centro de Astrof{\'{i}}sica y Tecnolog{\'{i}}as Afines ACE210002 (CATA). This research has made use of the SIMBAD database,
operated at CDS, Strasbourg, France. This work has been co-funded by the National Science Centre, Poland,
grant No. 2022/45/B/ST9/00243. 
\end{acknowledgements}

 \begin{appendix}

\section{Error estimates}

We estimate the errors of mass, radius and orbital separation assuming an uncertainty in the temperature and using an analytical approach. The luminosity of a follows the Stefan-Boltzmann law:

\begin{equation}
L = 4 \pi R^2 \sigma T^4
\end{equation}

\noindent
where R is the radius of the star, $\sigma$ is the Stefan-Boltzmann constant and $T$ is the effective temperature. Now, an error in the temperature, 
$\Delta T$, affects the luminosity due to the dependence $L$ $\propto$ $T^{4}$. If we make an error $\Delta T$ in measuring the effective temperature of the star, the relative error in the luminosity will be:

\begin{equation}
\frac{\Delta L}{L} = 4 \frac{\Delta T}{T} 
\end{equation}

Given that for a main sequence star $L \propto M^{\alpha}$, an error in the luminosity, $\Delta L$, translates into an error in the mass, $\Delta M$, as follows:

\begin{equation}
\frac{\Delta M}{M} = \frac{1}{\alpha} \frac{\Delta L}{L} = \frac{4}{\alpha} \frac{\Delta T}{T}
\end{equation}

If we use $\alpha \approx$ 3.5 for a main sequence star, and consider that for a giant star the exponent should be larger because of the increasing radius, the relative error in the mass of the donor would be:

\begin{equation}
\frac{\Delta M}{M} < \frac{4}{3.5} \frac{\Delta T}{T} \approx  1.14 \frac{\Delta T}{T} 
\end{equation}

On the other hand, the orbital separation in a binary system is given by Kepler's third law:

\begin{equation}
a = (\frac{G(M_1+M_2)P^2}{4\pi^2})^{1/3}
\end{equation}

An error in the stellar masses affects the orbital separation as follows:

\begin{equation}
\frac{\Delta a}{a} = \frac{1}{3} \frac{\Delta (M_1 + M_2)}{M_1 + M_2} 
\end{equation}

Assuming the relative error is similar for both stars, we can write:

\begin{equation}
\frac{\Delta a}{a} = \frac{1}{3} \frac{4}{\alpha} \frac{\Delta T}{T} 
\end{equation}

For $\alpha$ $>$ 3.5:

\begin{equation}
\frac{\Delta a}{a} <  \frac{1.14}{3} \frac{\Delta T}{T}  \approx 0.38    \frac{\Delta T}{T}
\end{equation}

From the above equations, and assuming a temperature uncertainty of 4\%, we get masses with uncertainty less than 5\% and orbital separation with uncertainty less than 2\%.

To calculate the error in the radius of the giant star filling its Roche lobe, we need to consider how the errors in the mass ratio (\(q = M_1 / M_2\)) and the orbital separation (\(a\)) affect the error in the Roche lobe radius (\(R_L\)).

Recall that the Roche lobe radius is given by the empirical Eggleton formula:

\begin{equation}
\frac{R_L}{a} = \frac{0.49 q^{2/3}}{0.6 q^{2/3} + \ln(1 + q^{1/3})}
\end{equation}


To find the error in \(R_L\) (the Roche lobe radius, which will approximately be the radius of the giant star), we will apply the formula for error propagation. This will allow us to calculate the error in \(R_L\) as a function of the errors in \(q\) and \(a\).

Since \(R_L\) depends on two variables \(q\) and \(a\), the error in \(R_L\), denoted as \(\Delta R_L\), can be computed using the following error propagation formula:

\begin{equation}
\frac{\Delta R_L}{R_L} = \sqrt{\left( \frac{\partial R_L}{\partial q} \cdot \frac{\Delta q}{q} \right)^2 + \left( \frac{\partial R_L}{\partial a} \cdot \frac{\Delta a}{a} \right)^2}
\end{equation}

First we calculate the partial derivative with respect to \(q\):

\tiny
   \begin{equation}
   \frac{\partial}{\partial q} \left( \frac{R_L}{a} \right) = \frac{0.49 \cdot \frac{2}{3} q^{-1/3}}{\left(0.6 q^{2/3} + \ln(1 + q^{1/3}) \right)} - \frac{0.49 q^{2/3} \cdot \left( 0.6 \cdot \frac{2}{3} q^{-1/3} + \frac{1}{3(1 + q^{1/3}) q^{2/3}} \right)}{\left(0.6 q^{2/3} + \ln(1 + q^{1/3})\right)^2}
   \end{equation}
\normalsize

Then,  we calculate the partial derivative with respect to \(a\). The dependence of \(R_L\) on \(a\) is straightforward, as \(R_L \propto a\). Thus:

   \begin{equation}
   \frac{\partial R_L}{\partial a} = \frac{R_L}{a}
   \end{equation}

For $q$ = 0.19 with an associated error of 0.02, and assuming a 4\% relative error in the orbital separation, we obtain a relative error of 3.54\% for the Roche Lobe radius, which also corresponds to the donor star's radius.

%
%
%

If the gainer follows a mass-radius relationship $M \propto R^{\beta}$ with $\beta$ between 0.57 and 0.7, we get the fractional error in the gainer radius less than 0.7  times the fractional error in the gainer mass. If this last is  5\%, then the gainer fractional radius uncertainty is less than 4\%.

Finally, the surface gravity in solar units is given by:

\begin{equation}
 \log g = \log \left( \frac{M}{R^2} \right) + \log g_{\odot}
   \end{equation}

\noindent
where $\log g_{\odot}$ is the solar surface gravity. The error is given by:

\begin{equation}
\Delta (\log g) = \sqrt{\left( \frac{\Delta M}{M \ln(10)} \right)^2 + \left( \frac{2 \Delta R}{R \ln(10)} \right)^2}
 \end{equation}

Using the fractional errors quoted above, we derive the errors for log\,g given in Table \ref{tab:system}.

\section{Additional material}

\begin{table*}
\centering
\caption{Epochs of main eclipses (HJD - 2450000).}
\label{tab:minimatimes}
\small
\begin{tabular}{rrrrr} 
\hline
Epoch&Epoch&Epoch&Epoch&Epoch\\
\hline
-1169.6891& 506.9281&4896.5155&5926.7238&8444.7825\\
-522.0847 &1119.0054&4896.5829&5927.7229&8445.6026\\
-407.8935 &2644.7548&4896.6069&5964.6214&8445.7437\\
-407.7107 &2682.7154&4934.4964&5965.6203&8445.8333\\
-406.9255 &2911.8491&4935.4862&6002.5849&9856.8030\\
-369.7454 &2950.6943&5850.7912&6003.5510&9856.8183\\
-368.9330 &3408.6229&5851.8002&7339.7611&9856.8330\\
-368.7462 &3446.5358&5888.7934&7453.6087&9856.8470\\
-331.8936 &4399.7700&5889.8355&8444.5994&9895.6433\\
432.0125  &4476.6967&5925.7121&8444.6805&9895.6574\\
\hline
9895.6717&9933.6693&9971.6538 &10009.5486   &10010.5759 \\
9895.6857&9933.7353&9971.6684&10009.5626&10010.5951\\
9895.7001&9933.7498&9971.6845&10009.5824&10010.6170\\
9895.7142&9971.5522&9971.6986&10009.5970&10010.6311\\
9895.7286&9971.5665&9971.7129&10009.6159&10010.6454\\
9932.7679&9971.5805&9971.7268&10009.6299&\\
9932.7961&9971.5948&9971.7411&10009.6444&\\
9933.5796&9971.6088&9971.7552&10010.5153&\\
9933.6407&9971.6235&10009.5153&10010.5296&\\
9933.6548&9971.6397&10009.5293&10010.5442&\\
\hline
\end{tabular}
\end{table*}

\begin{table*}
\centering
\caption{ Parameters of the light curve fits.}
\label{tab:fitpar}
\small
\begin{tabular}{ccrrrrrrrrrrrr}
\hline 
set&    $F_{d}$ & $T_{disc}$&   $A_{hs}$&  $\theta_{hs}$ & $\lambda_{hs}$ & $\theta_{rad}$ &  $A_{bs}$&  $\theta_{bs}$ & $\lambda_{bs}$ &  $a_T$   & $R_{disk}$ & $d_e$& $d_c$\\
& $\pm$0.01 & $\pm$60 & $\pm$0.04 & $\pm$1.2 & $\pm$ 3.0 & $\pm$3.0 & $\pm$0.04 & $\pm$1.3 &$\pm$ 8.0  & $\pm$0.02 & $\pm$0.1 & $\pm$0.1 & $\pm$0.2\\
 & &(K)&&(\dg)&(\dg)&(\dg)&&(\dg)&(\dg)&&(\rsun)&(\rsun)&(\rsun) \\
\hline
01 &0.886& 	3475& 	1.36& 	16.8& 	328.0& 	-5.1	& 1.13& 	44.4& 	101.8& 	0.74& 	42.99& 	6.68	& 3.77 \\
02 &0.898& 	3258& 	1.24& 	16.7& 	347.8& 	17.1	& 1.16& 	45.6& 	161.2& 	0.75& 	43.56& 	9.52	& 4.43 \\
03 &0.888& 	3432& 	1.27& 	20.3& 	345.2& 	-8.6	& 1.08& 	38.8& 	169.6& 	0.75& 	43.10& 	8.14	& 4.47   \\
04 &0.866& 	3418& 	1.21& 	19.3& 	337.5& 	-3.2	& 1.09& 	40.5& 	155.0& 	0.75& 	42.00& 	8.57	& 4.51    \\
05 &0.903& 	3391& 	1.23& 	22.1& 	341.2& 	3.3	& 1.15& 	39.3& 	162.3& 	0.74& 	43.81& 	9.09	& 4.64 \\
06 &0.749& 	3985& 	1.20& 	25.4& 	332.1& 	-2.8	& 1.12& 	34.7& 	93.7	 &    0.49& 	36.35& 	9.03	& 3.46 \\
07 &0.747& 	3745& 	1.13& 	21.9& 	348.0& 	16.9	& 1.09& 	32.8& 	59.7	 &    0.73& 	36.26& 	9.04	& 4.67   \\
08 &0.724& 	3678& 	1.07& 	23.5& 	349.1& 	25.5	& 1.07& 	36.9& 	61.3	 &    0.74& 	35.13&    10.21	& 5.02    \\
09 &0.839& 	3600& 	1.35& 	16.7& 	328.6& 	-2.3	& 1.14& 	48.4& 	81.4	 &    0.72& 	42.12& 	6.54	& 3.81 \\
10 &0.752& 	3729& 	1.30& 	17.5& 	340.0& 	5.1	& 1.19& 	48.1& 	77.1	 &    0.75& 	36.47& 	6.47	& 4.46 \\
11 &0.710& 	3802& 	1.33& 	18.1& 	333.2& 	6.5	& 1.21& 	48.2& 	75.1	 &    0.75& 	34.20& 	6.20	& 4.42   \\
12 &0.753& 	3664& 	1.26& 	14.7& 	316.3& 	-31.5	& 1.18& 	42.0& 	59.1	 &    0.74& 	36.55& 	7.06	& 4.53    \\
13 &0.762& 	3585& 	1.30& 	15.2& 	310.7& 	-2.3	& 1.16& 	41.8& 	55.2	 &    0.73& 	36.98& 	8.32	& 4.53 \\
14 &0.772& 	3739& 	1.39& 	19.7& 	310.5& 	12.9	& 1.18& 	49.9& 	92.7	 &    0.75& 	37.44&    10.73	& 4.54 \\
15 &0.726& 	3681& 	1.39& 	22.0& 	317.7& 	6.5	& 1.18& 	44.0& 	98.0	 &    0.75& 	35.20&    10.40	& 4.54   \\
16 &0.736& 	3648& 	1.40& 	20.8& 	322.2& 	-12.2	& 1.16& 	43.6& 	98.9	 &    0.74& 	35.71& 	9.83	& 4.45    \\
17 &0.872& 	3692& 	1.38& 	17.4& 	313.1& 	4.9	& 1.12& 	49.4& 	118.1&     0.72& 	42.32& 	6.40	& 3.61 \\
18 &0.881& 	3602& 	1.35& 	19.7& 	323.0& 	1.2	& 1.21& 	50.5& 	64.0	 &    0.72& 	42.73& 	6.07	& 3.06 \\
19 &0.865& 	3522& 	1.18& 	15.1& 	325.6& 	-13.6	& 1.15& 	47.4& 	66.9	 &    0.69& 	41.96& 	7.35	& 3.37   \\
20 &0.896& 	3545& 	1.34& 	16.9& 	328.5& 	-12	& 1.18& 	48.7& 	66.7	 &    0.75& 	43.48& 	6.89	& 3.75    \\
21 &0.756& 	3851& 	1.21& 	15.2& 	316.0& 	18.4	& 1.07& 	35.2& 	102.4&    0.70&         36.66& 	10.9	& 4.53 \\
22 &0.756& 	3863& 	1.27& 	15.9& 	318.5& 	22.8	& 1.08& 	36.1& 	106.1&    0.73&         36.70& 	9.76	& 4.51 \\
23 &0.729& 	3847& 	1.28& 	16.4& 	326.9& 	28.7	& 1.08& 	39.1& 	82.1	 &    0.74& 	35.36&    10.48	& 4.53   \\
24 &0.659& 	4217& 	1.30& 	15.9& 	310.6& 	11.3	& 1.1	0&    42.9& 	106.0&    0.72&         31.97&    12.86	& 4.09    \\
25 &0.677& 	4114& 	1.28& 	16.0& 	316.8& 	-9.2	& 1.15& 	47.9& 	79.6	 &    0.67& 	32.84& 	9.62	& 3.38 \\
26 &0.699& 	3846& 	1.39& 	18.2& 	321.9& 	1.1	& 1.16& 	50.4& 	91.7	 &    0.74& 	33.90& 	9.89	& 3.71 \\
27 &0.732& 	3902& 	1.21& 	25.1& 	326.5& 	-11.6	& 1.13& 	35.4& 	66.7	 &    0.69& 	35.54&    10.01	& 4.41   \\
\hline
Mean & 0.786	 & 3697& 1.282 & 18.6 & 327.2 & 2.5 & 1.138 & 43.0 & 94.5 & 0.72 & 38.20 & 8.74 & 4.19 \\
std  & 0.077       & 220    & 0.085 & 3.1 & 12.1 & 13.9 & 0.043 & 5.5 & 33.3 & 0.05 & 3.81 & 1.78 & 0.50  \\
\hline
\vspace{0.05cm}
\end{tabular}\\
Note: Light curve fit parameters  with their formal errors for the  datasets. Mean and standard deviation are also given. See text for details. 
\end{table*}

\end{appendix}
\end{document}